\documentclass{article}
\usepackage{graphicx}
\usepackage{amsmath}
\usepackage{amsfonts}
\usepackage{amssymb}
\newtheorem{theorem}{Theorem}

\newtheorem{definition}[theorem]{Definition}

\begin{document}

\title{Constructive proposals for QFT based on the crossing property and on
lightfront holography\\{\small Dedicated to Jacques Bros on the occasion of his 70th birthday}}
\author{Bert Schroer\\presently: CBPF, Rua Dr. Xavier Sigaud 150 \\22290-180 Rio de Janeiro, Brazil\\Prof. em., Institut f\"{u}r Theoretische Physik, FU-Berlin\\email: schroer@cbpf.br}
\date{May 26 2004}
\maketitle
\begin{abstract}
The recent concept of modular localization of wedge algebras suggests two
methods of classifying and constructing QFTs, one based on particle-like
generators of wedge algebras using on-shell concepts (S-matrix, formfactors.
crossing property) and the other using the off-shell simplification of
lightfront holography (chiral theories).

The lack of an operator interpretation of the crossing property is a serious
obstacle in on-shell constructions. In special cases one can define a
``masterfield'' whose connected formfactors constitute an auxiliary thermal
QFT for which the KMS cyclicity equation is identical to the crossing property
of the formfactors of the master field.

Further progress is expected to result from a conceptual understanding of the
role of on-shell concepts as particle states and the S-matrix within the
holographic lightfront projection.
\end{abstract}

\section{History of the crossing property}

The so-called crossing property of the S-matrix and formfactors\footnote{In
the setting of formfactors i.e. matrix elements of operators between
multiparticle ket in-states and bra out-states the S-matrix is a special case
of a (generalized) formfactor associated with the identity operator.} is a
deep and important, but at the same time incompletely understood structure in
particle physics. As a result of its inexorable link with analyticity
properties in the quantum field theoretic setting of scattering theory,
crossing is not a symmetry in the standard sense (of Wigner), even though it
is often referred to as ``crossing symmetry''.

In contrast to the underlying causality principles which are ``off-shell'',
i.e. are formulated in terms of local observables or fields with unrestricted
Fourier transforms, the crossing property is ``on-shell'', that is to say it
refers to particle states which are described by wave functions on the forward
mass hyperboloid $p^{2}=m^{2},p^{0}\geq0$. Particle properties are intrinsic
to a theory, whereas fields are (point-like \cite{S-W} or
string-like\cite{M-S-Y}\cite{Mu}\cite{prep}) ``coordinatizations'' of local
algebras; only local equivalence classes of fields or the local algebras
generated by fields are truely ``intrinsic''\footnote{The individuality of
classical fields is lost in QFT where e.g. a meson field is any local
(relative local with respect to the local obervables of the theory) covariant
object with a nonvanishing matrix element between the vacuum and a one-meson
state (``interpolating field'').}. The use of the notion of ``intrinsicness''
in local quantum physics (LQP) is reminiscent of the use of ``invariant'' (as
opposed to coordinate-dependent) in geometry; in this analogy the coordinates
in geometry correspond to the coordinatization of spacetime-indexed algebras
by pointlike field generators. More specifically, the use of pointlike fields
is analogous to the use of singular coordinates (coordinate systems which
become singular somewhere) since quantum fields are ``operator-valued
distributions'' which require smearing with test functions.

In the Lagrangian quantization approach to QFT, as well as in the more
intrinsic algebraic approach to LQP, crossing plays no significant role. Only
in formulations of particle physics which start with on-shell quantities and
aim at the construction of spacetime-indexed local algebras or local
equivalence classes of fields, the crossing becomes an important structural tool.

Examples par excellence of pure on-shell approaches are the various attempts
at S-matrix theories which aim at direct constructions of scattering data
without the use of local fields and local observables. The motivation behind
such attempts was for the first time spelled out by Heisenberg
\cite{Heisenberg} and amounts to the idea that by limiting oneself to
particles and their mass-shells, one avoids (integration over) fluctuation on
a scale of arbitrarily small spacelike distances which are the cause of
ultraviolet divergencies.

This idea of giving constructive prominence to ``on-shell'' aspects is quite
different and certainly more conservative than attempts at improving
short-distance properties by introducing non-local interactions in a field
theoretic framework (for a historical review of non-local attempts see
\cite{cluster}) which generally causes grave problems with the causality
properties underlying particle physics. The main purpose of approaches using
scattering concepts (``on-shell'') is to avoid such inherently singular
objects as pointlike fields in calculational steps, which is a reasonable aim
independent of whether one believes that a formulation of interactions in
terms of singular pointlike fields exists in the mathematical physics sense or not.

Heisenberg's S-matrix proposal can be seen as the first attempt in this
direction. It incorporated unitarity, Poincar\'{e} invariance and certain
analytic properties, but already run into problems with the implementation of
cluster factorization properties for the multiparticle scattering.

There exists a more recent scheme of ``direct particle interaction'' which
solved this cluster factorization problem for the multi-particle
representations of the Poincar\'{e} group in the presence of interactions by
an iterative construction \cite{C-P}. To understand the problem with
clustering, it is helpful to recall that in multiparticles Schr\"{o}dinger
quantum mechanics the step from n to n+1 particles by simply adding the
two-particle interactions of the new particle with the n previous ones
manifestly complies (for sufficiently short range interactions) with the
cluster factorizability of the unitary representors of the 10-parametric
Galilei-group; the system and its symmetries factorizes into previously
constructed subsystems. But this infinite ``Russian matrushka'' picture of
particle physics (iteratively adding particles together with their
interactions and in turn recovering the previous smaller systems by
translating one of the particle to infinity) runs into serious problems in the
relativistic context. In mathematical terms there exists a mismatch between
the adding-on of particles and their L-covariant interactions on the one hand,
and the cluster factorizability property i.e. the tensor factorization of the
representation into the representation of the previously encountered
multi-particle subsystems on the other hand. For the two-particle systems
there is no problem with clustering if one defines the interaction in terms of
an additive modification of the invariant two-particle mass operator as first
proposed by Bakamijan and Thomas \cite{B-T}. However the iteration of this B-T
procedure to 3 particles leads to a Poincar\'{e} covariant representation
which fails to cluster (the Hamiltonian and the L-boosts are not
asymptotically additive); although the 3-particle S-matrix\footnote{The
possibility of two-particle bound states entering as incoming particles
requires the use of the framework of rearrangement collisions in which the
space of (noninteracting) fragments is distinguished from the (Heisenberg)
space on which the interacting Poincar\'{e} group is represented \cite{C-P}.}
does cluster \cite{Coester}. Adding a fourth particle in the B-T way would
also lead to the breakdown of the 4-particle S-matrix clustering. The solution
to this obstruction was later found in \cite{C-P}; it consisted in modifying
the 3-particle system by adding a connected 3-particle interaction in such a
way that the 3-particle S-matrix does not change. This is done by a so-called
``scattering equivalence''\cite{Sok} i.e. a unitary transformation which
changes the (Bakamijan-Thomas) 3-particle representation without affecting the
3-particle S-matrix\footnote{Whereas in QFT the permitted field changes which
maintain the local net of algebras are described by the local equivalence
classes (Borchers classes), the scattering equivalences in the C-P scheme form
a much bigger nonlocal class of changes.}.

It turns out that this process of adding on interactions to the mass operator
and then enforcing clustering by invoking scattering equivalence works
iteratively \cite{C-P} and yields an n-particle interacting representation of
the Poincar\'{e} group; in particular one obtains M\o ller operators and an
S-matrix which fulfill the cluster factorization property. There is a prize to
pay, namely the use of scattering equivalences prevent the use of a second
quantization formalism known from Schr\"{o}dinger QM, thus separating
relativistic direct particle interactions from QFT even on a formal level.
Nevertheless it does secure the macro-locality expressed by the (rapid in case
of short range interactions) fall-off properties of the connected parts of the
representation of the Poincar\'{e} group and the S-matrix. Different from the
mass superselection rule in Galilei invariant quantum mechanics, there is no
selection rule involving particle masses which requires the absence of
particle creation processes coming from Poincar\'{e} symmetry in this
relativistic direct particle interaction formalism \cite{C-P}. This poses the
interesting question whether by coupling channels which lead to an increasing
number of created particles one can approximate field theoretic models by
mathematically controllable direct particle interactions. After this interlude
about the feasibility of macro-causal relativistic particle theory (for a more
detailed presentation see \cite{cluster}) we now return to the setting of QFT.

Since the early 1950s, in the aftermath of renormalization theory, the
relation between particles and fields received significant elucidation through
the derivation of time-dependent scattering theory. It also became clear that
Heisenberg's S-matrix proposal had to be amended by the addition of the
crossing property i.e. a prescription of how to analytically continue particle
momenta on the complex mass shell in order to relate matrix elements of local
operators between incoming ket and outgoing bra states with a fixed total sum
of in + out particles in terms of one ``masterfunction''. In physical terms it
allows to relate matrix elements with particles in both the incoming ket- and
outgoing bra-states to the vacuum polarization matrix elements where the
ket-state (or the bra state) is the vacuum vector.

Whereas Heisenberg's requirements on a relativistic S-matrix can be
implemented in a direct particle interaction scheme, the implementation of
crossing is conceptually related to the presence of vacuum polarization for
which QFT with its micro-causality is the natural arena. At this point it
should be clear to the reader why we highlighted the little known direct
particle interaction theory; if one wants to shed some light on the mysterious
crossing symmetry, it may be helpful to contrast it with theories of
relativistic particle scattering in which this property is absent.

The LSZ time-dependent scattering theory and the associated reduction
formalism relates such a matrix element (referred to as a generalized
formfactor) in a natural way to one in which an incoming particle becomes
``crossed'' into an antiparticle on the backward real mass shell; it is at
this point where analytic continuation from a physical process enters. The
important remark here is that the use of particle states requires the
restriction of the analytic continuation to the complex mass shell
(``on-shell''). If one were to allow off-shell analytic continuations, the
derivation of the crossing would be much easier since it would then follow
from off-shell spectral representations of the Jost-Lehmann-Dyson kind or
perturbatively from Feynman diagrams and time-ordered functions. In this paper
the notion of crossing will only be used in the restrictive on-shell analytic
continuation as it is needed for on-shell relation between formfactors.

A rigorous on-shell derivation for two-particle scattering amplitude has been
given by Bros\footnote{Since the issue of crossing constitutes the main
subject of the BEG paper, I find it particularly appropriate to dedicate this
work to Jacques Bros on the occasion of his 70$^{th}$ birthday.}, Epstein and
Glaser \cite{Br-Ep}. The S-matrix is the formfactor of the identity operator.
In the special case of the elastic scattering amplitude, the crossing of only
one particle from the incoming state has to be accompanied by a reverse
crossing of one of the outgoing particles in order to arrive at a physical
process allowed by energy-momentum conservation. This crossing of a pair of
particles from the in/out elastic configuration is actually the origin of the
terminology ``crossing'' and was the main object of rigorous analytic
investigations \cite{Br-Ep}. A derivation of crossing in the setting of QFT
for general multi-particle scattering configurations and for formfactors, as
one needs it for the derivation of a bootstrap-formfactor program (see later)
from the general principles of local quantum physics, does not yet exist. It
is not clear to me whether the present state of art in QFT would permit to go
significantly beyond the old and still impressive results quoted before
\cite{Br-Ep}.

The crossing property became the cornerstone of the so-called bootstrap
S-matrix program and several ad hoc representations of analytic scattering
amplitudes were proposed (Mandelstam, Regge...) in order to incorporate
crossing in a more manageable form.

An interesting early historical chance to approach QFT from a different
direction by using on-shell global objects without short distance
singularities was wasted when the S-matrix bootstrap approach ended in a
verbal cleansing rage against QFT\footnote{One glance at the old conference
proceedings and review articles of the Chew S-matrix school reveals that I am
not exaggerating. Nowadays the ideological fervor against QFT is hard to
understand, in particular in view of the fact that almost all the concepts
originated from QFT.} instead of serving in its construction as attempted in
this paper.

Some of the S-matrix boot strap ideas were later used by Veneziano \cite{Ve}
in the construction of the ``dual model''. But there is an essential
difference in the way crossing was implemented. Whereas the field theoretic
crossing involves a finite number of particles with the scattering continuum
participating in an essential way, the dual model implements crossing without
the continuum by using instead as a start discrete infinite ``particle tower''
with ever increasing masses (the origin of what was later called
``stringyness''). This tower structure was afterwards interpreted in terms of
the particle excitations of a relativistic string. It is important to note
that Veneziano's successful mathematical experiment to implement crossing with
properties of Gamma functions was more than a mathematical invention. In the
late 60s there some of the dominant phenomenological ideas about Regge poles
called for a one-particle ``saturation'' of the crossing property in the
setting of Mandelstam's representation of the 2-particle scattering amplitude.
The popularity which the dual model enjoyed before QCD appeared on the scene
was more related to these phenomenological aspects rather then to its role in
carrying some of the legacy of the S-matrix bootstrap approach.

There is some irony in the fact that Chew and his followers, who tried to find
a philosophical basis for their S-matrix bootstrap ideas to attain the status
of a theory of everything (TOE), did not succeed in these
attempts\footnote{The S-matrix bootstrap returned many years later as a
valuable tool (but not a TOE) of the ``formfactor program'' in the limited
context of d=1+1 factorizing models of QFT\cite{K-W}.}, whereas Veneziano, who
had no such aims, laid the seeds of string theory. Contrary to the original
phenomenological intentions of the dual model, its string theoretical
re-interpretation elevated it in the eyes of some physicist to the status of a
TOE (this time including gravity). Whether one subscribes to such view or not,
there can be little doubt that string theory became quite speculative and
acquired a somewhat ideological stance. Contrary to the bootstrap of the Chew
school however, it led to significant mathematical enrichments even though its
role for particle physics became increasingly mysterious.

The main reason why the old bootstrap approach ended in the dustbin of history
was its clinging to its dismissive view of QFT even at a time when the success
of gauge theories was already obvious. On a deeper level and in and in
relation to the content of the present paper it is obvious that it did not
succeed in its own terms since it was unable convert the analyticity based
bootstrap ideas by a mathematically well-defined operator formalism which
incorporates the crossing property in a natural way.

In recent years the similarity of the cyclic crossing property of formfactors
with the better understood cyclic KMS condition for wedge-localized algebras
(the Rindler Unruh thermal aspect) led to the conjecture that the former is an
on-shell consequence of the latter. Whereas this turns out to be true for
d=1+1 factorizing models, the nature of the connection between these two
cyclic properties in the general setting remains obscure and needs further clarifications.

The content of the paper is organized as follows. In the next section we set
the stage for the concept of modular localization which will be our main new
constructive tool. Whereas without interactions there is a complete
parallelism between particle- and field- modular localization, the presence of
interactions has a de-localizing effect on the side of particles as a result
of interaction-caused vacuum polarization. A useful concept which captures
this de-localization aspect is that of vacuum-\textbf{p}olarization-\textbf{f}%
ree \textbf{g}enerators (PFG) which highlights the wedge localization as
representing the best compromise between particle- and field- localization. In
the third section we recall that the requirement of translation invariant
domains for PFGs (``tempered'' PFGs) essentially leads to the
Zamolodchikov-Faddeev algebra structure which characterize d=1+1 factorizing
theories. This is a modest realization of the old ``bootstrap dream'', but now
as a valuable constructive tool of QFT without the unfounded claim of a TOE.

In the fourth section the idea of a ``masterfield'' will be set forth whose
connected formfactors define a nonlocal QFT in momentum space for which the
KMS condition is identical with crossing. Whereas for factorizing models this
idea reduces to Lukyanov's ``free field representations'', in a more general
setting the hypothesis remains a matter of interesting speculation and a
subject for future research.

Finally modular localization is used to formulate ``algebraic lightfront
holography'' which relates massive quantum field theories to generalized
chiral models on the lightfront. As a result of its firm anchoring in AQFT and
is conceptual tightness, one would expect this new idea to play an important
role in future construction methods. Its confrontation with the setting for
d=1+1 factorizing models reveals that the massive particle aspects including
crossing and scattering data and the chiral conformal field based holographic
properties coexist as two descriptions of the same theory in one and the same
Hilbert space.

\section{Modular Localization for Particles and Fields}

The concept of modular localization, which will be reviewed in this section,
has significantly enriched ideas about the relation between particles and
fields. In particular it has led to a profound understanding of those
properties in the particle-field relation which persist in the presence of
interactions and which in turn are important in an intrinsic understanding of
interaction; this is the understanding which, borrowing an aphorism of Pascual
Jordan \cite{Jordan}, does not rely on ``classical crutches'', as does the
standard Lagrangian quantization.

Historically the first step into a direction of intrinsic formulation of
relativistic quantum physics was undertaken by Wigner when in 1939 he
identified relativistic particle states with irreducible positive energy
representations of the Poincar\'{e} group. These representations come with two
localization concepts: the Newton-Wigner localization \cite{N-W} and the more
recent modular localization \cite{B-G-L}\cite{Mund}\cite{F-S}.

The N-W localization is the result of the adaptation of Born's quantum
mechanical localization probability density to Wigner's relativistic setting.
This localization is important in relativistic scattering theory since it
leads to the probability interpretation of cross sections, which was actually
the setting in which Born introduced probabilities into QM (the x-space
probability interpretation of the Schroedinger wave function appears later in
Pauli's Handbuch article). It is not Lorentz-covariant nor local\footnote{Far
from being a a peculiar shortcoming of the Newton-Wigner localization, there
exist a general No-Go theorem which rules out the existence of any
Poincar\'{e}-covariant localization in terms of projectors and probabilities
in theories with positive energy \cite{Mal}.} for finite distances, but the
fact that it acquires these two properties in the asymptotic region is
sufficient for obtaining a relativistic asymptotic particle description and in
particular a Poincar\'{e} invariant S-matrix \cite{Haag}. It should not come
as a surprise that its use for propagation over finite distances leads to
nonsensical results on the feasibility of superluminal propagation
\cite{Bu-Yng}.

On the other hand the modular localization is the localization which is
implicit in the formalism of local quantum field theory. It is well known that
if one applies smeared fields with localized $\mathcal{O}$-support of the
smearing function $suppf\subset\mathcal{O}$ to the vacuum, the resulting
vectors will belong to a dense subspace $H(\mathcal{O})\footnote{The denseness
of this subspace is the main content of the Reeh-Schlieder theorem
\cite{Haag}.}$ which will change its position in the ambient space with the
change of the localization region
\begin{equation}
A(f)\Omega\in H(\mathcal{O})\subseteq H
\end{equation}

Modular localization theory is a relatively new conceptual framework which
places this kind of relation between spacetime regions of vacuum excitations
and positions of dense subspaces on a more intrinsic and rigorous footing, so
that it becomes independent of the use of field coordinatizations. This is
done by trading the subspace generated by smeared fields with the domain of
the $\mathcal{O}$-dependent Tomita S-operator $H$($\mathcal{O}$)$\equiv
domS_{\mathcal{O}}$ (see next section) which is directly associated with the
localized algebra and does not refer to its coordinatization in terms of fields.

This encoding of Minkowski spacetime localization into relative position of
subspaces (or equivalently in terms of real subspaces (\ref{real}) of which
$H$($\mathcal{O}$) turns out to be the complex combination) is a
characteristic phenomenon of local quantum physics; it essentially depends on
the presence of a finite maximal causal propagation speed and hence has no
counterpart in the Schr\"{o}dinger QM. The denseness of the localization
spaces prevents a description in terms of projectors onto complex subspaces
and hence evades the assumptions of the mentioned no-go theorem \cite{Mal}.

This unusual situation, which goes somewhat against quantum mechanical
intuition, is inexorably linked with a structural change of the local algebras
as compared to the algebraic structure of quantum mechanics. Whereas the
algebra of QM has minimal projectors (corresponding to best observations), the
structure of projection operators within local relativistic algebras is very
different from that of projectors in the global algebra associated with the
entire Minkowski spacetime. All these changes can be traced back to the
omnipresence of vacuum polarizations which in turn are inexorably related to
relativistic causality in the setting of quantum theories.

The difference between quantum mechanical and modular localization shows up in
a dramatic fashion in a famous Gedankenexperiment which Fermi proposed
\cite{Fermi} in order to show that the velocity of light remains the limiting
propagation velocity in the quantum setting of relativistic field theory. An
updated argument confirming Fermi's conclusion which takes into account the
conceptual progress on the issue of causal localization and mathematical rigor
can be found, as mentioned before, in \cite{Bu-Yng}. Although all quantum
mechanical situations associated with Bell's inequalities can be transferred
to QFT with the help of the split property, there are problems with achieving
the vacuum polarization free two-particle state postulated by Fermi
\footnote{I am indebted to Larry Landau for reminding me of the problems of
using the split property in connection with the realization of the Fermi
Gedankenexperiment in the relativistic setting.}. This does however not affect
the conclusion that localized exitations of the vacuum cannot propagate with a
superluminal speed.

The modular localization theory associated with localized algebras in QFT has
a simpler spatial counterpart which can be directly applied to the Wigner
representation theory of the Poincar\'{e} group. In the next section we will
study this spatial modular localization. In addition of being interesting in
its own right, this will facilitate the subsequent presentation of algebraic
modular localization theory which is indispensable in order to incorporate
interactions in a field-coordinatization independent way.

\subsection{Modular localization in the absence of interactions}

Modular localization as an intrinsic concept of local quantum physics (i.e.
without reference to any pointlike field coordinatization), has its origin in
the Bisognano-Wichmann theorem for wedge-localized algebras in QFT
\cite{Bi-Wi}\cite{Bor}. In the context of Wigner's description of elementary
relativistic systems in terms of irreducible positive energy representations
of the Poincar\'{e} group, the construction of this localization proceeds as
follows \cite{B-G-L} \cite{Mund}\cite{F-S}

\begin{enumerate}
\item  Fix a reference wedge region, e.g. $W_{R}=\left\{  x\in\mathbb{R}%
^{4};x^{1}>\left|  x^{0}\right|  \right\}  $ and use the Wigner representation
of the $W_{R}$-affiliated boost group $\Lambda_{W_{R}}(\chi)$ and the
$x^{0}-x^{1}-$reflection\footnote{In certain cases the irreducible
representation has to be doubled in order to accomodate the antiunitary (time
is inverted) reflection. This is always the case with zero mass finite
helicity representations and more generally if particles are not
selfconjugate.} along the edge of the wedge $j_{W_{R}}$ in order to define the
following antilinear unbounded closable operator (with $closS=clos\Delta
^{\frac{1}{2}}$). Retaining the same notation for the closed operators, one
defines
\begin{align}
S_{W_{R}}  &  :=J_{W_{R}}\Delta^{\frac{1}{2}}\\
J_{W_{R}}  &  :=U(j_{W_{R}}),\;\Delta^{it}:=U(\Lambda_{W_{R}}(2\pi
t))\nonumber
\end{align}
The commutativity of $J_{W_{R}}$ with $\Delta^{it}$ together with the
antiunitarity of $J_{W_{R}}$ yield the property which characterize a Tomita
operator\footnote{Operators with this property are the corner stones of the
Tomita-Takesaki modular theory \cite{Takesaki} of operator algebras. Here they
arise in the spatial Rieffel van Daele spatial setting of modular theory from
a realization of the geometric Bisognano-Wichmann situation within the Wigner
representation theory.} $S_{W_{R}}^{2}\subset1$ whose domain is identical to
its range$.$ Such operators are well-known to be equivalent to their real
standard subspaces of the Wigner representation space $H$ which arise as their
closed real +1 eigenspaces $K(W)$%
\begin{align}
K(W_{R})  &  :=\left\{  \psi\in H,\,S_{W_{R}}\psi=\psi\right\} \label{real}\\
\overline{K(W_{R})+iK(W_{R})}  &  =H,\,\,K(W_{R})\cap iK(W_{R})=0\nonumber\\
J_{R}K(W_{R})  &  =K(W_{R})^{\bot}\nonumber
\end{align}
The real subspace $K(W_{R})$ is closed in $H$, whereas the complex subspace
spanned together with the -1 eigenspace $iK(W_{R})$ is the dense domain of the
Tomita operator $S_{W_{R}}$ and forms a Hilbert space in the graph norm of
$S_{W_{R}}$. The denseness in $H$ of this span $K(W_{R})+iK(W_{R})$ and the
absence of nontrivial vectors in the intersection $K(W_{R})\cap iK(W_{R})$ is
called ``standardness''. The right hand side in the third line refers to the
symplectic complement i.e. a kind of ``orthogonality'' in the sense of the
symplectic form $Im(\cdot,\cdot).$
\end{enumerate}

\textit{Additional comments}. The denseness of the complex spans of modular
localization spaces is a one-particle analog of the Reeh-Schlieder theorem
\cite{Haag}. Each Tomita operator $S_{W}$ encodes physical information about
localization into the position of its dense domain (which equals its dense
range) within $H.$ Equivalently real standard subspaces or their complex dense
span determine uniquely an abstract Tomita operator (which in general is not
related to geometry or group representation theory). The application of
Poincar\'{e} transformations to the reference situation generates a consistent
family of wedge spaces $K(W)=U(\Lambda,a)K(W_{R})$ if $W=(\Lambda,a)W_{R}.$

One of the surprises of this modular localization setting is the fact that it
already preempts the spin-statistics connection on the level of one-particle
representation theory by producing a mismatch between the symplectic and the
geometric complement which is related to the spin-statistics factor
\cite{Mund}\cite{F-S}
\begin{align}
K(W)^{\bot}  &  =ZK(W^{\prime})\\
Z^{2}  &  =e^{2\pi is}\nonumber
\end{align}
Another surprising fact is that the modular setting prepares the ground for
the crossing property, since the equation characterizing the real modular
localization subspaces in more details reads
\begin{equation}
\left(  J\Delta^{\frac{1}{2}}\psi\right)  (p)=\Sigma\overline{\psi_{c}%
(-p)}=\psi(p)
\end{equation}
i.e. the complex conjugate of the analytically continued wave function (but
now referring to the charge-conjugate situation$)$ is up to a matrix $\Sigma$
which acts on the spin indices equal to the original wave function.

\begin{enumerate}
\item [2]The sharpening of localization is obtained by intersecting wedges in
order to obtain real subspaces as causally closed subwedge regions:
\begin{equation}
K(\mathcal{O}):=\cap_{W\supset\mathcal{O}}K(W) \label{int}%
\end{equation}
The crucial question is whether they are ``standard''. According to an
important theorem of Brunetti, Guido and Longo \cite{B-G-L} standardness holds
for spacelike cones $\mathcal{O}=\mathcal{C}$ \ in all positive energy
representations. In case of finite spin/helicity representations the
standardness also holds for (arbitrary small) double cones $D$. The double
cone regions $D$ are conveniently envisaged as intersections of a forward cone
with a backward cone whose apex is inside the forward cone; the simplest
description of a spacelike cone $C$\ with apex $a$ is in terms of a scaled up
double cone $C=a+\cup_{\lambda\geq0}\lambda D$ where $D$ is spacelike
separated from the origin. Both regions are characteristic for simply
connected Poincar\'{e}-invariant causally closed families of compact or
noncompact extension resulting from intersecting wedges in Minkowski
spacetime. In those cases where the double cone localized spaces with
pointlike ''cores'' are trivial (massless infinite spin, massive d=1+2
\ anyons), the smallest localization regions are spacelike cones with
semiinfinite strings as cores.
\end{enumerate}

\textit{Additional comments.} Although the connection between standard real
subspaces and Tomita operators $S$ holds in both directions (and hence
standard intersections always have an associated Tomita operator $S),$ the
components of their polar decomposition $\Delta^{it}$ and $J$ have generally
no relations to diffeomorphisms of the underlying spacetime. While leaving the
localization regions invariant (or transforming them into their causal
disjoint) and hence still encoding the full information of localization, their
actions within $\mathcal{O}$ as well on its causal complement $\mathcal{O}%
^{\prime}$ are ``fuzzy'', which at best may be expressed (in the Wightman
setting of QFT) in terms of actions on test function spaces with fixed
localization supports (see below \ref{test}).

\begin{enumerate}
\item [3]In the absence of interactions the transition from free particles to
algebras of fields is most appropriately done in a functorial way by applying
the Weyl (CCR) (or in case of halfinteger spin the CAR functor) to the
localization K-spaces\footnote{To maintain simplicity we limit our
presentation to the bosonic situation and refer to \cite{Mund}\cite{F-S} for
the general treatment.}:
\begin{align}
\mathcal{A(O)} &  :=alg\left\{  Weyl(\psi)|\,\,\psi\in K(\mathcal{O})\right\}
\label{alg}\\
Weyl(f) &  :=expi\left\{  \int a^{\ast}(p,s)\psi(p,s)\frac{d^{3}p}{2\omega
(p)}+h.a.\right\}  \nonumber
\end{align}
The functorial relation between real subspaces and von Neumann algebras
preserves the causal localization structure \cite{L-R-T} and commutes with the
improvement of localization through intersections (\ref{int}) (denoted by
$\cap$) as expressed in the following commuting square
\end{enumerate}%

\begin{equation}%
\begin{array}
[c]{ccc}%
K_{W} & \longrightarrow & \mathcal{A}(W)\\
\downarrow{\small \cap} &  & \downarrow{\small \cap}\\
K_{\mathcal{O}} & \longrightarrow & \mathcal{A(O)}%
\end{array}
\label{square}%
\end{equation}
i.e. without interactions there is a perfect match between particle- and
field- localization\footnote{We retain the traditional word ``field'' in the
sense of carriers of causal localization even though the present construction
avoids the explicit use of pointlike operator-valued distributions.}. For
later purposes we introduce the following definition \cite{S2}.

\begin{definition}
A vacuum-\textbf{p}olarization-\textbf{f}ree \textbf{g}enerator (PFG) for a
region $\mathcal{O}$ is an operator affiliated with the algebra $\mathcal{A}%
$($\mathcal{O}$) which created a vacuum-polarization-free one-particle vector
\begin{align}
&  G\,\eta\text{ }\mathcal{A(O)}\\
G\Omega &  =1-particle\nonumber
\end{align}
\end{definition}

It is easy to see that (in case of Bosons) PFGs are necessarily unbounded
operators. In the absence of interactions they turn out to consist precisely
of those $\mathcal{O}$-localized operators which are linear in the Wigner
creation/annihilation operators. In that case a denumerable covariant
pointlike basis of PFGs is conveniently described in terms of the well-known
set of interwining functions $u(p,s)$ (and their charge conjugates $v(p,s)$)
which relate the given canonical $(m,s)$ Wigner representation with the
various tensorial (spinorial) covariant free fields
\begin{align}
A(x)  & =\int\left\{  e^{-ipx}\sum u(p,s_{3})a(p,s_{3})+e^{ipx}\sum
v(p,s_{3})b^{\ast}(p,s_{3})\right\}  \frac{d^{3}p}{2p^{0}}\label{field}\\
\,\;\,p^{0}  & =\sqrt{\vec{p}^{2}+m^{2}}%
\end{align}
Whereas the $(m,s)$ Wigner creation/annihilation operators $a^{\#}(p,s)$ and
the above localized algebras are unique, there exists an denumerable set
(labeled by pairs of undotted/dotted spinorial indices) of covariant
intertwinwers for fixed $(m,s)$ \cite{Wei}. Their main role with respect to
the issue of modular localization consists in relating the quantum concept of
modular localization to the more classical notion of localization via support
properties of test functions
\begin{equation}
K(\mathcal{O})=clos\left\{  E_{m}\tilde{f}(p)u_{k}(p,s)|\,suppf\subset
\mathcal{O},k=1...N\right\}  \label{test}%
\end{equation}
where $E_{m}f(p)$ stands for the mass-shell projection\footnote{As a result of
the mass shell restriction a Wigner wave function (and the smeared fields) is
represented in terms of an equivalence class of test functions.} of the
Fourier transform of the real test function $f$ and the closure is taken in
the linear span with $i$ runs over all Lorentz (spinorial) components $N$ and
$f$ running over all $\mathcal{O}$-supported test functions; as before the
closure within the Wigner representation space is restricted to real linear
combinations. This way of relating modular localization to classical test
function supports is (whenever it is possible) the easiest way to show the
standardness property. When the appearance of massless infinite spin
representations only allows standardness of spacelike cone-localized spaces,
the analogs of the above intertwiners lead to semiinfinite spacelike
string-localized fields $A(x,e)$ (with $e$ being a spacelike unit vector
\cite{M-S-Y}) which have no interpretation in terms of Lagrangian quantization
(and should not be confused with objects of string theory).

As expected, the crossing relation for connected matrix elements (connected
formfactors) of a wedge-localized operator $B\in\mathcal{A}(W)$ ($\bar{p}$
denotes the charge conjugate particle with momentum $p)$%
\begin{align}
&  \left\langle p_{1},...,p_{k}\left|  B\right|  p_{k+1},...,p_{n}%
\right\rangle _{conn}\label{cross}\\
&  =\left\langle -\bar{p}_{n},p_{k},...,p_{1}\left|  B\right|  p_{k+1}%
,...,p_{n-1}\right\rangle _{conn}\nonumber
\end{align}
results from the KMS property of the wedge-restricted vacuum state
($suppf_{i}\subset W$)
\begin{align}
&  \left\langle A(f_{1})^{\ast}...A(f_{k})^{\ast}BA(f_{k+1})...A(f_{n}%
)\right\rangle \\
&  =\left\langle Ad\Delta(A(f_{n}))A(f_{1})^{\ast}...A(f_{k})^{\ast}%
BA(f_{k+1})...A(f_{n-1})\right\rangle \nonumber
\end{align}
by taking the connected part and using the density of the W-supported product
of test functions in the multiparticle tensor-product Wigner spaces.

Since it is very convenient to consider the later lightfront holography
(section 6) as part of modular wedge localization, we will briefly explain in
the following in a pedestrian way how this is done for a massive Hermitian
free field $A(x)^{\ast}=A(x)$. Using the previous notation (\ref{test}) one
has for real test functions with $suppf\subset W$
\begin{align}
A(f) &  =\int\left(  a^{\ast}(p)E_{m}\tilde{f}(p)+h.c.\right)  \frac{d^{3}%
p}{2p_{0}}=\int\left(  a^{\ast}(p)E_{m}\tilde{f}(p)+h.c.\right)  \frac
{d\theta}{2}d^{2}p_{\perp}\\
\,\,\,\, &  \left[  a(p),a^{\ast}(p^{\prime})\right]  =2p^{0}\delta(\vec
{p}-\vec{p}^{\prime})=2\delta(\theta-\theta^{\prime})\delta(p_{\perp}%
-p_{\perp}^{\prime})\nonumber\\
&  with\text{ }p=(m_{eff}\cosh\theta,m_{eff}\sinh\theta,p_{\perp}%
),\,m_{eff}=\sqrt{m^{2}+p_{\perp}^{2}}\nonumber
\end{align}
where the $x^{0}-x^{1}$ localization in the $0$-$1$ reference wedge implies
that $E_{m}\tilde{f}(p)$ of the real test function $f$ is a vector in the
dense subspace $K_{r}(W)+iK_{r}(W)$ of boundary value of analytic functions in
the $\theta-$strip with respect to the measure $\frac{d\theta}{2}d^{2}%
p_{\perp}.$ Since product functions $E_{m}f(p)=\tilde{f}_{+}(\theta)\tilde
{f}_{\perp}(p_{\perp})$ with $\tilde{f}_{+}(\theta)$ strip-analytic are dense
in $K_{r}(W)+iK_{r}(W)$ it is convenient to use them in the following way
($p_{-}\equiv e^{\theta}$)%
\begin{align}
&  \int\left(  a^{\ast}(p)\tilde{f}_{+}(\theta)\tilde{f}_{\perp}(p_{\perp
})+h.c.\right)  \frac{d\theta}{2}d^{2}p_{\perp}=A(f_{+}f_{\perp})\\
&  =\int A_{LF}(x)f_{+}(x_{+})f_{\perp}(x_{\perp})dx_{+}dx_{\perp}\nonumber\\
&  f_{+}(x_{+})\equiv\frac{1}{2\pi}\int_{0}^{\infty}\tilde{f}_{+}(\ln
p_{-})e^{ip_{-}x_{+}}\frac{dp_{-}}{2p_{-}}\;\nonumber\\
&  A_{LF}(x)=\frac{1}{\left(  2\pi\right)  ^{\frac{3}{2}}}\int\left(  a^{\ast
}(p)e^{ip_{-}x_{+}+ip_{\perp}x_{\perp}}+h.c.\right)  dp_{-}d^{2}p_{\perp
}\nonumber\\
&  \left[  a(p),a^{\ast}(p^{\prime})\right]  =2p_{-}\delta(p_{-}-p_{-}%
^{\prime})\delta(p_{\perp}-p_{\perp}^{\prime})\;\nonumber\\
&  \curvearrowright\left\langle A_{LF}(x)A_{LF}(x^{\prime})\right\rangle =\int
e^{-ip_{-}(x_{+}-x_{+}^{\prime})}\frac{dp_{-}}{2p_{-}}\cdot\delta(x_{\perp
}-x_{\perp}^{\prime})\nonumber
\end{align}
where in the last line the two-point function has been rewritten in the new
lightfront variables. As a consequence of the strip analyticity in $\theta$
the function $f_{+}(x_{+})$ is supported on the positive $x_{+}$ axis. Note
that the vanishing of the Fourier transform at $p_{-}=0$ is not imposed but
results from the square integrability of $\tilde{f}_{+}(\theta)$ which forces
the $\tilde{f}_{+}(\ln p_{-})$ to vanish at the lower boundary $p_{-}=0$ (this
also holds without the specialization to product functions)$.$

Without this vanishing property the infrared divergence in the Fourier
representation for $A_{LF}(x)$ would not be compensated and the expression
would not be equal to the original one. The relevant testfunction spaces for
lightcone quantization were first introduced (without referring to modular
localization) in \cite{Driessler}. Note also that the Fourier transformed
lightfront test functions $f_{+}(x_{+})f_{\perp}(x_{\perp})$ (unlike their
original counterpart $f(x))$ are not subject to any mass shell restriction
i.e. the lightfront localization relates the smeared fields with individual
functions on the lightfront rather than mass shell equivalence
classes\footnote{We are referring to the fact that the relation between
testfunctions $f$ and their wave functions $E_{m}\tilde{f}$ in the Wigner
one-particle space is an equivalence class relation.} of ambient test functions..

The terminology ``lightfront restriction'' for this rewriting becomes more
comprehensible in terms of the following formal steps ($r=\sqrt{x_{1}%
^{2}-x_{0}^{2}}$)
\begin{align}
&  A(x)|_{W}=A(r\sinh\chi,r\cosh\chi,x_{\perp})\nonumber\\
&  =\frac{1}{\left(  2\pi\right)  ^{\frac{3}{2}}}\int\left(  a^{\ast
}(p)e^{im_{eff}r\sinh(\chi-\theta)+ip_{\perp}x_{\perp}}+h.c.\right)
\frac{d\theta}{2}d^{2}p_{\perp}\\
&  \overset{\chi=\ln r,r\rightarrow0}{\longrightarrow}\,\frac{1}{\left(
2\pi\right)  ^{\frac{3}{2}}}\int\left(  a^{\ast}(p)e^{im_{eff}e^{\theta}%
x_{+}+ip_{\perp}x_{\perp}}+h.c.\right)  \frac{d\theta}{2}d^{2}p_{\perp
}\label{lim}\\
&  =\frac{1}{\left(  2\pi\right)  ^{\frac{3}{2}}}\int d^{2}p_{\perp}\int
_{0}^{\infty}\left(  a^{\ast}(p)e^{ip_{-}x_{+}+ip_{\perp}x_{\perp}%
}+h.c.\right)  \frac{dp_{-}}{2p_{-}}=A_{LF}(x)=:A(x)|_{LF}\nonumber
\end{align}
where in the last line we have absorbed $m_{eff}$ into the definition of the
integration variable $p_{-}.$ Although we obtain the same formula as before,
the formal way requires to add the restriction on test function spaces whose
Fourier transforms vanish at $p_{-}=0$ ``by hand''. For d=1+1 the transverse
$x_{\perp}$ and $p_{\perp}$ are absent.

Lightfront restriction does not mean pointwise restriction of the correlation
functions i.e. $\left\langle A(x)A(x^{\prime})\right\rangle _{x_{-}%
=0=x_{-}^{\prime}}\neq\left\langle A(x)A(x^{\prime})\right\rangle |_{LF}%
\equiv\left\langle A|_{LF}(x)A|_{LF}(x^{\prime})\right\rangle $. This point
was the source of occasional confusion in the literature on lightcone
quantization. In fact already the terminology ``lightcone quantization''
creates the impression that one is aiming at a different quantization leading
to a possibly different theory, whereas in reality the physical problem is to
describe the ambient local theory in terms of a different locality structure
associated with the lightfront. This LF locality structure, although being
local in its own right, is relatively nonlocal with respect to the ambient
locality structure. The pivotal problem of how these two structures are
related was not addressed in the old approach.

In the absense of interactions the lightfront restriction $A|_{LF}$ shares
with the ambient free field $A$ the vanishing of higher than two-point
correlations. As a consequence there is only one ambient theory associated
with the above lightfront field. As will be argued in section 6, in the
presence of interactions one expects the relation of the ambient theories to
their holographic projection to be many to one i.e. the concept of
``holographic universality classes'' becomes important in inverse holography
(reconstruction of ambient theories from a given LF description).

The important observation in the context of localization is that the algebras
generated by smearing $A(x)|_{W}$ and $A(x)|_{LF}$ with the corresponding test
function spaces are identical%
\begin{equation}
alg\left\{  A(f)\;|\;suppf\subset W\right\}  =alg\left\{  A_{LF}(f_{+}%
f_{\perp})\;|\;suppf_{+}\subset\mathbb{R}_{+}\right\}
\end{equation}

Although the equality of the wedge- with the lightfront- localized algebra
turns out to be a general feature of QFT\footnote{The only exception is the
case of massless theories in d=1+1.}, it is only in the free field case that
one can describe the localization aspects of the lightfront algebra by the
above process of a restriction of the ambient free field. For interacting
fields the local net structure on the lightfront has to be recovered in an
algebraic manner referred to as ``algebraic lightfront holography'', which
will be presented in section 6. This new approach demystifies and corrects to
a considerable degree the old ideas on lightcone quantization.

\subsection{Modular localization in the presence of interactions}

There is a drastic weakening in the relation between particle- field
localization when interactions are present. The parallelism expressed in the
above commuting square is lost. In particular interactions destroy the
possibility of having subwedge-localized PFGs\footnote{The J-S theorem can
easily be generalized to subwedge-localized PFGs \cite{Mu}.
\par
{}}. Quantum fields also loose that kind of ``individuality''(associated with
the measurement of field strength) which fields enjoy in classical physics;
the role of quantum fields (besides being the non-intrinsic implementers of
the relativistic locality principle) is restricted to interpolate particles
and to ``coordinatize'' (in the sense of singular generators) local nets of
algebras. Hence it is somewhat surprising that there are two remarkable and
potentially useful properties which survive the presence of interactions. As
in the framework of LSZ scattering theory, in the following we are assuming
the existence of a mass gap.

\begin{enumerate}
\item  Wedge algebras $\mathcal{A}(W)$ have the smallest localization region
which still permits affiliated PFGs \cite{B-B-S}, i.e. to every
wedge-localized one-particle wave function $\psi\in K(W)+iK(W)$ there exists a
$G_{\psi}\eta\mathcal{A}(W)$ with
\begin{align}
G_{\psi}\Omega &  =\psi\\
G_{\psi}^{\ast}\Omega &  =S\psi\nonumber
\end{align}
This is the best compromise between particles and fields in the presence of
interactions; any improvement on the level of particles (e.g. construction of
n-particle states for n%
%TCIMACRO{\TEXTsymbol{>}}%
%BeginExpansion
$>$%
%EndExpansion
1) would only be possible in the completely de-localized global algebra (which
contains e.g. the creation/annihilation operators). Vice versa any improvement
in the localization by passing to subwedge algebras would lead to the
admixture of interaction-induced vacuum polarization (states with ill-defined
particle number) to the one-particle component. Hence the presence of this
kind of vacuum polarization clouds for subwedge regions is an intrinsic signal
of the presence of interactions. This raises the interesting question whether
there is some common feature to interaction-induced vacuum polarization clouds
which permits a finer classification of interactions; this is a problem which
certainly must be solved if one wants to use this intrinsic characterization
of interactions as a constructive alternative to the more extrinsic
field-coordinatization dependent standard Lagrangian quantization approach.

\item  In asymptotically complete QFT, the S-matrix $S_{scat}$ is a relative
modular invariant between the interacting and the free incoming wedge
algebras
\begin{align}
S  &  =J\Delta^{\frac{1}{2}}\\
\Delta^{it}  &  =\Delta_{in}^{it},\,\,J=J_{in}S_{scat}\nonumber
\end{align}
This follows from the TCP-invariance of the S-matrix and the fact that the
modular $J$ differs from TCP by a spatial $\pi$-rotation \cite{Haag} which (as
all connected Poincar\'{e} transformations) commutes with the scattering
matrix. This structural property relates the position of the dense
wedge-localized subspace $H_{F}(W)$ within the Fock space $H_{F}$ (defined by
e.g. the out-operators) to the S-matrix.

\item  The split property \cite{Haag} permits to formulate the notion of
``statistical independence'' (well-known from quantum mechanics) which
concerns the construction of interacting states with independently prescribed
local components. This is needed in order to control the strong vacuum
fluctuations which result from sharp spacetime localization and leads to a
partial return of quantum mechanical structures. Although the split property
has up to now not played a direct role in model constructions, it is believed
to be important in securing the standardness of intersections of wedge
algebras and hence the nontriviality of models \cite{Bu-Le}.
\end{enumerate}

\textit{Additional comments. }The interpretation of the scattering operator as
a relative modular invariant associated with the wedge region leads to rather
strong consequences if one assumes that the connected part of the formfactors
fulfill the following crossing relations%

\begin{align}
&  ^{out}\left\langle p_{1},...,p_{k}\left|  B\right|  p_{k+1},...,p_{n}%
\right\rangle _{conn}^{in}=\label{crossing}\\
&  ^{out}\left\langle -\bar{p}_{n},p_{k},...,p_{1}\left|  B\right|
p_{k+1},...,p_{n-1}\right\rangle _{conn}^{in}\nonumber
\end{align}
where $B$ is an operator affiliated with $\mathcal{A}(W).$ It has the same
form as in the free case (\ref{cross}) except that the particles in the
bra/ket vectors are referring to the different out/in particle states.
Evidently this property permits to relate the vacuum polarization components
\begin{equation}
\left\langle p_{n},...,p_{1}|B\Omega\right\rangle \label{comp}%
\end{equation}
with the general formfactor by a succession of crossings. The position of the
dense subspace generated by all operators $B\eta\mathcal{A}(W)$ affiliated
with $\mathcal{A}(W)$ from the vacuum is determined by the domain of the
Tomita operator $S$ which is in turn determined by the scattering operator
$S_{scat}.$ Assume that a given crossing symmetric scattering operator
$S_{scat}$ would admit two different wedge algebras $\mathcal{A}%
_{i}(W),\,i=1,2.$ Since these algebras must have the same Tomita operator for
each $B_{1}\eta\mathcal{A}_{1}(W)$ there must exist an operator $B_{2}%
\eta\mathcal{A}_{1}(W)$ such that $B_{1}\Omega=B_{2}\Omega$ which means that
the vacuum polarization components (\ref{comp}) are identical$.$ But then the
crossing property (\ref{crossing}) lift this identity to the general
formfactors which requires $B_{1}=B_{2}$ and hence the desired equality
$\mathcal{A}_{1}(W)=\mathcal{A}_{2}(W).$ Since the net of localized algebras
is uniquely fixed in terms of intersections of wedge algebras, this would
imply the uniqueness of the inverse scattering problem \cite{inverse}. Note
however that the crossing property of formfactors in the general interacting
case is presently an additional assumption\footnote{The assumption of crossing
for formfactors as one needs it for the uniqueness of inverse scattering seems
to go beyond what has derived by the analyticity techniques in \cite{Br-Ep},
but a definite conclusion on this matter can probably not obtained without
updating these old but still impressive methods.}; only for d=1+1 factorizing
models crossing it can be shown to follow from the KMS property for the
restriction of the vacuum to wedge algebras in a similar fashion as for free
fields (see section 4). Without assuming the crossing property for formfactors
it does not appear to be possible to derive the uniqueness of the inverse
scattering problem from the standard postulates of QFT \cite{Bu-Fr}.

The prerequisites for formfactor crossing are obtained from the LSZ scattering
theory and in particular from the resulting reduction formulas in terms of
time-ordered products. For the connected formfactors one obtains%

\begin{align}
&  ^{out}\left\langle q_{1},q_{2},...q_{m}\left|  B\right|  p_{n}%
,...p_{2},p_{1}\right\rangle _{conn}^{in}=\label{red}\\
&  -i\int\,^{out}\left\langle q_{2},...q_{m}\left|  K_{y}TBA^{\ast}(y)\right|
p_{1},p_{2}...p_{n}\right\rangle _{conn}^{in}d^{4}ye^{-iq_{1}y}\nonumber\\
&  =-i\int\,^{out}\left\langle q_{1},q_{2},...q_{m}\left|  K_{y}TBA(y)\right|
p_{2}...p_{n}\right\rangle _{conn}^{in}d^{4}ye^{ip_{1}y}\,\nonumber
\end{align}
Here the time-ordering $T$ involving the original operator $B\in
\mathcal{A}(\mathcal{O})$ \ and the pointlike interpolating Heisenberg
field\footnote{The notion of interpolating fields and associated reduction
formulas cease to exist if the in/out particles require the application of
semiinfinite string-like Heisenberg operators to the vacuum.} $A(x).$ The
latter appears in the reduction of a particle from the bra- or ket state. For
the definition of the time ordering between a fixed finitely localized
operator $B$ and a field with variable localization $y$ we may use
$TBA(y)=\theta(-y)BA(y)+\theta(y)A(y)B,$ however as we place the momenta
on-shell, the definition of time ordering for $y$ near $locB$ fortunately
turns out to be irrelevant\footnote{For far separated $y$ we may consider
loc$B$ to be near zero; then $\theta(y)\approx\theta(y-locB)$ agrees
approximately with the relative timelike distance $\theta$-function used for
pointlike localization.}. These on-shell reduction formulas remain valid if
one used as interpolating operators instead of pointlike fields the translates
of bounded compactly localized operators \cite{Araki}. Each such reduction is
accompanied by another disconnected contribution in which the creation
operator of an outgoing particle $a_{out}^{\ast}(q_{1})$ changes to an
incoming annihilation $a_{in}(q_{1})$ acting on the incoming configuration;
there is a corresponding contraction term if we would reduce a particle from
the incoming state vector. These disconnected terms (which contain formfactors
with two particle less in the bra- and ket- vectors) have been omitted since
they do not contribute to generic nonoverlapping momentum contributions and to
the analytic continuations (and hence do not enter the connected part).

Under the assumption that there is an analytic path from $p\rightarrow
-p\,$\ (or $\theta\rightarrow\theta-i\pi,p_{\perp}\rightarrow-p_{\perp}$ in
the rapidity parametrization of the standard wedge), the comparison between
the two expressions gives the desired crossing property that is to say a
particle of momentum p in the incoming ket state within the formfactor is
crossed into an outgoing bra antiparticle at the analytically continued
momentum -p (here denoted as -\={p}) and the connected and the connected
formfactor remains invariant.

Reduction formulas and the crossing property are characteristic for pointlike
localized fields (corresponding to double cone localization in the algebraic
setting), their derivation breaks down \cite{BFloc} if interacting fields only
permit string-like localization (corresponding to the singular limit of
spacelike cone localization). The reason for this is that it is not enough to
control the localization of endpoints but one also must take care of the
spacelike string direction; but the kinematical requirement for having
convergence to outgoing asymptotic multi-particle states is different from
that for incoming states so that there exist no single interpolating field
which converges in both asymptotic directions. The particle-field relation and
the constructions derived from it exclude string-localized fields. However
this does not necessarily exclude string theory since there is no indication
that string theory is string-localized (see also the concluding remarks).

In order to obtain an analytic path on the complex mass-shell for e.g. the
$2\rightarrow2$ scattering amplitude it is convenient to pass from time
ordering $T$ to retardation $R$
\begin{equation}
TBA(y)=RBA(y)+\left\{  B,A(y)\right\}
\end{equation}
The unordered (anticommutator) term does not have the pole structure on which
the Klein-Gordon operator $K_{y}$ can have a nontrivial on-shell action and
therefore drops out. The application of the JLD spectral representation puts
the p-dependence into the denominator of the integrand of an integral
representation from where the construction of an analytic path interpolating
the formfactors with its crossed counterpart proceeds in an analog fashion to
the derivation of crossing for the S-matrix \cite{BLOT}\cite{Araki}. Whereas
it is fairly easy to find an off-shell analytic path, the construction of an
on-shell path i.e. one which remains in the complex mass shell is a
significantly more difficult matter \cite{Br-Ep}. The LSZ reduction formalism
is suggestive of crossing but for themselves too weak for securing the
mathematical existence of paths on the complex mass shell which link real
forward and backward mass shells.

The simplifications of the LSZ formalism resulting from factorizability of
models can be found in an appendix of \cite{BFKZ}

The result of the comparison between the reduction (\ref{red}) applied to
outgoing and incoming configurations may be written in the following
suggestive way (for spinless particles)%

\begin{align}
&  ^{out}\left\langle p_{1},p_{2},...p_{l}\left|  B\right|  q_{k}%
,_{k-1}...q_{2},q_{1}\right\rangle ^{in}=\label{cro}\\
&  \underset{q_{c}\rightarrow-q_{1}}{a.c.}^{out}\left\langle q_{c},p_{1}%
,p_{2},...p_{l}\left|  B\right|  q_{k},q_{k-1}...q_{2}\right\rangle
^{in}+c.t\nonumber\\
&  =:^{out}\left\langle -q_{1},p_{1},p_{2},...p_{l}\left|  B\right|
q_{k},q_{k-1}...q_{2}\right\rangle ^{in}+c.t.\nonumber
\end{align}
where the contraction terms c.t. involve momentum space $\delta$-functions
(which are part of the LSZ reduction theory) and the last line denotes a
shorthand notation for the analytic continuation to the real negative mass
shell. Instead of crossing from incoming ket to outgoing bras one may of
course also cross in the reverse direction from bras to kets. The important
physical role of the crossing property is to relate the vacuum polarization
components of an operator to the connected part of the transition it causes
between in and out scattering states via iterated crossing
\begin{equation}
^{out}\left\langle p_{1},p_{2},...p_{n}\left|  B\right|  \Omega\right\rangle
\overset{iteration}{\longrightarrow}\,^{out}\left\langle p_{k},p_{k+1}%
,...p_{n}\left|  B\right|  -\bar{p}_{k-1}...-\bar{p}_{2},-\bar{p}%
_{1}\right\rangle _{conn}^{in}%
\end{equation}
Note that the vacuum polarization components are always connected. It is very
important to realize that the simplicity of the crossing property occurs only
for the connected part of the matrix elements; in order to write down the
relation for the full matrix elements one must keep track of all the momentum
space contraction terms in the iterative application of the LSZ formalism. It
is the connected part which is described by one analytic ``master
function''\ whose different boundary values correspond to the connected part
of the different matrix elements. This already indicates that one should
expect problems if one wants to understand crossing as an operational property
in the original theory of operators since taking connected parts of
correlation functions is not expressible as an operator algebraic property.
Indeed attempts to relate crossing to the cyclicity property of thermal
expectation values in KMS states on operator algebras within the general
setting of QFT failed\footnote{The structural similarity between the cyclicity
of the crossing- with the KMS-property has lured many authors (including the
present author \cite{Modloc}) into conjectures that crossing has a KMS
interpretation in the setting of wedge-localized algebras of the original
theory. These conjectures (including ``proofs'' \cite{Nieder}) are incorrect.}.

.

\section{The bootstrap-formfactor program in d=1+1 factorizing QFT}

As mentioned in the introduction, a modest but in its own right very
successful version of the S-matrix bootstrap with strong field theoretic roots
emerged in the second half of the 70s from some prior quasiclassical
integrability discoveries \cite{Dashen}. These seemingly exact quasiclassical
observations on the special two-dimensional as the ''Sine-Gordon'' model of
QFT required an explanation beyond quasiclassical approximations \cite{S1}.
This line of research led finally to a general program of a
bootstrap-formfactor construction of so-called d=1+1 factorizable models
\cite{K-W}\cite{Zam}\cite{Smir}. From this new nonperturbative scheme for
constructing a particular class of field theories came a steady flux of new
models and it continues to be an important innovative area of research. Our
interest in the present setting lies in the potential messages it contains
with respect to a mass-shell based constructive approach without the
``classical crutches'' which underlie the Lagrangian quantization approach. In
particular we are interested in a better understanding of formfactor crossing.

This formfactor program uses the very ambitious original S-matrix bootstrap
idea in the limited context of a d=1+1 S-matrix Ansatz in which $S$ factorizes
into 2-particle elastic components $S^{(2)}$. A consequence of this
simplification is that the classification and calculation of factorizing
S-matrices \cite{KTTW} can be separated from the problem from the construction
of the associated off-shell QFT. Hence the S-matrix bootstrap becomes the
first step in a bootstrap-formfactor program, followed by a second step which
consists in calculating generalized formfactors of fields and operators beyond
the identity operator (which represents the S-matrix). One does not expect
such a two-step approach to be possible beyond factorizable models, rather the
construction of the S-matrix (which may be considered as the special
formfactor of the identity operator between in-out multi-particle states) is
expected to have to be carried out as part of the formfactor construction.

It is interesting to note that the calculated formfactors of those factorizing
models which possess continuously varying coupling parameters turn out to be
analytic functions (below the threshold of formation of bound states) with a
finite radius of analyticity around zero coupling strength \cite{BFKZ}%
\cite{Ba-Ka}. For the correlation function on the other hand one does not
expect expandability into a power series since their perturbative structure is
not visibly different from that of other strictly renormalizable models and
there exist general arguments against the convergence of perturbative series.
This raises the interesting question of whether such a dichotomy between
perturbatively converging on-shell objects versus nonconverging (at best
asymptotic) series for off-shell correlation functions may continue to hold in
general. It would be quite startling if on-shell quantities as formfactors in
renormalizable field theories have improved perturbative convergence
properties which are not shared by correlation functions.

The main motivating idea in favor of an on-shell approach, namely the total
avoidance of ultraviolet divergences, is convincingly vindicated in the
setting of factorizing models. The pointlike fields, which in the present
state of development of factorizing models are only known via their
multi-particle formfactors \cite{Ba-Ka}, have an interesting
interaction-induced vacuum structure in that they possess no PFGs localized in
subwedge regions. In other words despite their lack of real particle creation
through scattering processes, they nevertheless have the full vacuum
polarization structure which one expects in an interacting QFT and which in
turn is the prerequisite for the appearance of interaction-caused anomalous
short distance dimensions. In this respect of short distance behavior
factorizing models are more realistic than the (non-factorizing) polynomial
interactions in d=1+1 whose complete mathematical control was achieved with
the methods of ``constructive QFT'' \cite{G-J} (see also\cite{Jae} \ for
recent applications in a more algebraic QFT setting). It seems that the ``hard
analysis'' methods of the constructivists are restricted to
superrenormalizable models whose short distance behavior is not worse than
that of free fields, whereas presently the modular methods, which avoid using
singular field coordinatizations altogether, work best for factorizing models.

Contrary to the cluster property and macro-causality which, as we have seen,
are also implemented in the relativistic particle-based theory of ``direct
particle interactions'' \cite{C-P}), crossing is the characteristic imprint
which relativistic micro-causality leaves in on-shell restrictions of QFT.
Although the on-shell aspects of d=1+1 factorizing models appear at first
sight associated with a kind of one-dimensional relativistic
particle-conserving quantum mechanics (due to the absence of real particle
creation via scattering), a closer look reveals a significant difference which
already makes itself felt on the level of the particle-conserving S-matrix.
Its crossing property leads to a bound state picture which has become known
under the name ``nuclear democracy'' as opposed to the quantum mechanical
hierarchy with respect to the issue of bound versus elementary issue. Nuclear
democracy is the statement that in interacting QFT all stable one-particle
one-particle states are on the same footing apart from their superselected
charges which are subject to hierarchical fusion laws. The hierarchy is that
between fundamental and composite (fused) superselected charges, whereas the
particles are the asymptotically stable carries of unconfined charges.

If, as e.g. in the case of the Sine-Gordon model, one still misses operators
which carry fundamental charges which cannot be obtained by fusion (but rather
permit to represent the charges of the known particles as fused fundamental
charges), then the representation theoretical approach of the superselection
theory in the setting of AQFT reconstructs the missing charges and particles.
The reason why the presence of the latter is easily overlooked in the standard
formalism is that these more fundamental particles do not appear directly, but
only manifest themselves through particle-antiparticle vacuum polarization
``clouds'' in intermediate states of correlation functions. The theory of
superselection sectors extends the original theory in such a way that these
new charges and their possible particle carriers are naturally incorporated so
that their scattering can be described in terms of interpolating fields. It is
the principle of locality which permits the construction of full-fledged field
algebras from observable algebras and arrive in this way at a fundamental
understanding of the concept of internal symmetries as a consequence of the
local representation theory of observable nets of operator algebras
\cite{Haag}.

The fact that the bootstrap-formfactor approach to factorizable models does
not need special prescriptions, but that its ``axioms'' \cite{Smir} follow
from general principles of QFT becomes particularly transparent if the
construction is placed into the setting of Tomita-Takesaki modular theory of
operator algebras as adapted to the local quantum physics setting (also
referred to as the method of modular localization) \cite{S2}\cite{B-B-S}%
\cite{Lech}. This will be illustrated in some detail in the next section.

This setting also highlights the ``existence problem of QFT'' in a new and
promising fashion \cite{Bu-Le}\cite{Lech2}. Here we remind the reader that
even after almost eight decades after its discovery, and despite impressive
perturbative and asymptotic successes, the description of interacting
particles by covariant fields in 4-dimensional Minkowski spacetime remained
part of mathematically as well as conceptually uncharted territory. This
applies in particular to the ``standard model'' which is a source of a very
specific permanent discomfort unknown in other areas of theoretical physics.
The predictive success of this model, if anything, highlights the seriousness
of this problem which without that success would be of a more academic nature.

The algebraic basis of the bootstrap-formfactor program for the special family
of d=1+1 factorizable theories is the validity of a momentum space
Zamolodchikov-Faddeev algebra \cite{Z}. The operators of this algebra are
close to free fields in the sense that their Fourier transforms are on-shell
(see \ref{PFG} in next section) objects, but they are non-local in the
pointlike sense. A closer look reveals that they are localizable in the weaker
sense of generating wedge algebras\footnote{An operator which is localizable
in a certain causally closed spacetime region is automatically localized in
any larger region but not necessarily in a smaller region. The unspecific
terminology ``non-local'' in the literature is used for any non pointlike
localized field.} \cite{S2}\cite{Lech}. In fact the existence of ``tempered''
(existence of a well-defined Fourier transform) wedge localized PFGs which
implies the absence of real particle creation through scattering processes
\cite{B-B-S} turns out to be the prerequisite for the success of the
bootstrap-formfactor program for factorizable models in which one uses only
formfactors and avoids (short-distance singular) correlation functions.

According to an old structural theorem which is based on certain analytic
properties of a field theoretic S-matrix \cite{Aks}\cite{B-B-S}, virtual
particle creation without real particle creation is only possible in d=1+1
theories. This in principle leaves the possibility for direct 3- or higher-
particle elastic processes beyond two particle scattering. An argument by
Karowski (private communication) based on formfactor crossing shows that this
is inconsistent with the absence of real particle creation. In this sense the
Z-F algebra structure, which is at the heart of factorizing models, turns out
to be a consequence of special properties of PFG for modular
wedge-localization, a fact which places the position of the factorizing models
within QFT into sharper focus. The crossing property is encoded into the
two-particle scattering amplitude from where it is subsequently passed on to
the formfactors. In line with the previous unicity argument of inverse
scattering based on crossing, the bootstrap formfactor approach associates
precisely one local equivalence class of fields (one net of localized operator
algebras) to a factorizing S-matrix. It also goes a long way in securing the
existence of operators whose matrix elements in multi-particle states give
rise to these explicitly computed formfactors.

In agreement with the philosophy underlying AQFT, which views pointlike fields
as coordinatizations of generators of localized algebras, the
bootstrap-formfactor construction for d=1+1 factorizing models primarily aims
to determine coordinatization-independent double-cone algebras by computing
intersections of wedge algebras. The nontriviality of a theory is then
tantamount to the nontriviality ($\neq\mathbb{C}\mathbf{1}$) of such
intersections\footnote{See a recent review \cite{Cherno} in which the minimal
formfactor contributions, which are a joint property of the local equivalence
class, have been separated from the polynomial contributions (the
``p-functions'') which distinguish between the vacuum polarization
contributions of individual fields.}. The computation of a basis of pointlike
field generators of these algebras is analogous but more involved than the
construction of the basis of composites of \ free fields which are the Wick
polynomials. Even for noninteracting theories the functorial description of
the algebras (\ref{alg}) based on modular localization is conceptually simpler
than the use of free fields (\ref{field}) and their local equivalence class of
Wick-ordered composites.

The crossing property is the crucial property which links scattering data with
off-shell operators spaces. As explained in the previous section, it relates
the multiparticle component of vectors obtained by one-time application of a
local (at least wedge-localized) operator to the vacuum with the connected
formfactors of this operator. It is importanr to note that in factorizing
models crossing is not an assumption but rather follows from the properties of
tempered PFGs for wedge algebras.

It is not easy to think of a formfactor approach beyond factorizing models. We
will present an operational idea of crossing which in principle does not
suffer from the above limitations of temperate PFGs, although one is presently
only able to test it in the d=1+1 factorizing setting. It is based on the
working hypothesis that each quantum field theory possesses a distinguished
field called a ``masterfield'' whose connected parts of its formfactors
defines a global (i.e. no local substructure) quantum field theory in the
on-shell momentum space variables. This auxiliary theory is in a thermal state
at the KMS Hawking temperature in such a way that the cyclic KMS property (the
thermal aspect of modular theory) is identical with the cyclic crossing
property. By construction this theory obeys momentum space cluster
decomposition properties in the rapidity variables. The simplicity of d=1+1
factorizing models finds its expression in the fact that the auxiliary
operator, whose KMS correlation functions are identified with the connected
formfactors of the masterfield, is an exponential of a bilinear expression in
free creation/annihilation operators. There is a good chance that this
structure is characteristic for factorizing models.

The subsequent content of the paper is organized as follows. The next section
recalls some details about the role of the Zamolodchikov-Faddeev algebra in
the generation of the modular wedge-localized operator algebra. After that we
will present two ideas which could be important in modular localization-based
constructions without assuming factorizability. One of these ideas consists in
postulating the already mentioned ``masterfield'' which generalizes
observations on cluster properties in momentum-rapidity space \cite{Cherno} as
well as observations on ``free field representations'' of formfactors in
factorizing models \cite{Lukyanov}. Another less speculative idea is to
classify and construct theories from their holographic lightfront projections,
which will be the subject of the last section before we present some conclusions.

\section{The Zamolodchikov-Faddeev algebra and its relation to modular localization}

In this section we recall some details about how the modular localization
formalism supports the bootstrap-formfactor construction.

It has been my Leitmotiv for a number of years \cite{Modloc} that the spirit
behind Wigner's representation theoretical approach enriched with the concept
of modular localization (as presented in the second section) could lead to a
truly intrinsic constructive approach in QFT which avoids those classical
quantization crutches which already the protagonist of field quantization
Pascual Jordan wanted to overcome. It was natural to test this idea first in
models which are similar to free field models in that their wedge-localized
algebras can be generated by fields which posses on-shell Fourier transforms.

In the previous section we learned that this class is related with the
Zamolodchikov-Faddeev algebra structure. In the simplest case of a scalar
chargeless particle without bound states\footnote{A situation which in case of
factorizing models with variable coupling (as e.g. the Sine-Gordon theory) can
always be obtained by choosing a sufficiently small coupling. Bound state
poles in the physical $\theta$-strip require nontrivial changes (e.g. the
$\phi$-generator is only wedge localized on the subspace of $Z$-particles)
which will be dealt with eleswhere.} the wedge generators are of the form \cite{S2}%

\begin{align}
\phi(x)  &  =\frac{1}{\sqrt{2\pi}}\int(e^{ip(\theta)x(\chi)}Z(\theta
)+h.c.)d\theta\label{PFG}\\
Z(\theta)Z^{\ast}(\theta^{\prime})  &  =S^{(2)}(\theta-\theta^{\prime}%
)Z^{\ast}(\theta^{\prime})Z(\theta)+\delta(\theta-\theta^{\prime})\nonumber\\
Z(\theta)Z(\theta^{\prime})  &  =S^{(2)}(\theta^{\prime}-\theta)Z(\theta
^{\prime})Z(\theta)\nonumber
\end{align}
Here $p(\theta)=m(ch\theta,sh\theta)$ is the rapidity parametrizations of the
d=1+1 mass-shell and $x=r(sh\chi,ch\chi)$ parametrizes the right hand wedge in
Minkowski spacetime; $S^{(2)}(\theta)$ is a structure function of the Z-F
algebra which is a nonlocal $^{\ast}$-algebra generalization of canonical
creation/annihilation operators. The notation preempts the fact that
$S^{(2)}(\theta)$ is the analytic continuation of the physical two-particle
S-matrix $S^{(2)}(\left|  \theta\right|  )$ which via the factorization
formula determines the general scattering operator $S_{scat}$ (\ref{fac}). The
unitarity and crossing of $S_{scat}$ follows from the corresponding
two-particle properties which in terms of the analytic continuation are
$S^{2}(z)^{\ast}=S^{(2)}(-z)$ (unitarity) and $S^{(2)}(z)=S^{(2)}(i\pi-z)$
(crossing) \cite{KTTW}. The $Z^{\ast}(\theta)$ operators applied to the vacuum
in the natural order $\theta_{1}>\theta_{2}>...>\theta_{n}$ are by definition
equal to the outgoing canonical Fock space creation operators whereas the
re-ordering from any other ordering has to be calculated according to the Z-F
commutation relations e.g.
\begin{equation}
Z^{\ast}(\theta)a^{\ast}(\theta_{1})a^{\ast}(\theta_{2})...a^{\ast}(\theta
_{n})\Omega=\prod_{i=1}^{k}S^{(2)}(\theta-\theta_{i})a^{\ast}(\theta
_{1})a^{\ast}(\theta_{2})..a^{\ast}(\theta).a^{\ast}(\theta_{n})\Omega
\end{equation}
where $\theta<\theta_{i}$ $i=1..k,\,\theta>\theta_{i}$ $i=k+1,..n.$ The
general Zamolodchikov-Faddeev algebra is a matrix generalization of this structure.

It is important not to identify the Fourier transform of the momentum with a
localization variable. Although the $x$ in $\phi(x)$ behaves covariantly under
Poincar\'{e} transformations, it is not marking a causal localization point;
in fact it is non-local variable in the sense of the standard use of this
terminology\footnote{The world local is reserved for ``commuting for spacelike
distances''. In this work we are dealing with non-local fields which are
nevertheless localized in causally complete subregions (wedges, double cones)
of Minkowski spacetime.}. It is however wedge-localized in the sense that the
generating family of operator for the right-hand wedge $W$ Wightman-like
(polynomial) algebra $alg\left\{  \phi(f),suppf\subset W\right\}  $ commutes
with the TCP transformed algebra $alg\left\{  J\phi(g)J,suppg\subset
W\right\}  $ which is the left wedge algebra \cite{Lech}
\begin{align}
&  \left[  \phi(f),J\phi(g)J\right]  =0\\
J &  =J_{0}S_{scat}\nonumber
\end{align}
Here$\ J_{0}$ is the TCP symmetry of the free field theory associated with
$a^{\#}(\theta)$ and $S_{scat}$ is the factorizing S-matrix which on
(outgoing) n-particle states has the form
\begin{equation}
S_{scat}a^{\ast}(\theta_{1})a^{\ast}(\theta_{2})...a^{\ast}(\theta_{n}%
)\Omega=\prod_{i<j}S^{(2)}(\theta_{i}-\theta_{j})a^{\ast}(\theta
_{2})...a^{\ast}(\theta_{n})\Omega\label{fac}%
\end{equation}
if we identify the $a^{\#}(\theta)$ with the incoming creation/annihilation
operators. It is then possible to give a rigorous proof \cite{Lech} that the
Weyl-like algebra generated by exponential unitaries is really wedge-localized
and fulfills the Bisognano-Wichmann property
\begin{align}
\mathcal{A}(W) &  =alg\left\{  e^{i\phi(f)}\,|\,\;suppf\subset W\right\}
\label{BW}\\
\mathcal{A}(W)^{\prime} &  =J\mathcal{A}(W)J=\mathcal{A}(W^{\prime})\nonumber
\end{align}
where the dash on operator algebras is the standard notation for their von
Neumann commutant and the dash on spacetime regions stands for the causal
complement. Within the modular setting the relative position of the causally
disjoint $A(W^{\prime})$ depends via $S_{scat}$ on the dynamics. The operator
TCP operator $J$ is the (antiunitary) angular part of the polar decomposition
of Tomita's algebraically defined unbounded antilinear S-operator with the
following characterization
\begin{align}
SA\Omega &  =A\Omega,\,\,A\in\mathcal{A}(W)\\
S &  =J\Delta^{\frac{1}{2}},\,\,\Delta^{it}=U(\Lambda(-2\pi t))\nonumber
\end{align}
with $\Lambda(\chi)$ being the Lorentz boost at the rapidity $\chi.$

At this point the setup looks like relativistic quantum mechanics since the
$\phi(f)$ (similar to genuine free fields if applied to the vacuum) do not
generate vacuum polarization clouds. The advantage of the algebraic modular
localization setting is that vacuum polarization is generated by algebraic
intersections which is in agreement with the intrinsic definition of the
notion of interaction presented in terms of PFGs in the previous section
\begin{align}
\mathcal{A}(D) &  \equiv\mathcal{A}(W)\cap\mathcal{A}(W_{a}^{\prime
})=\mathcal{A}(W)\cap\mathcal{A}(W_{a})^{\prime}\,\label{rel}\\
D &  =W\cap W_{a}^{\prime}\nonumber
\end{align}
This is the operator algebra associated with a double cone $D$ (which is
chosen symmetric around the origin by intersecting suitably translated wedges
and their causal complements). Note the difference from the quantization
approach, where pointlike localized fields are used from the outset and the
sharpening of localization of smeared products of fields is simply achieved by
the classical step of restricting the spacetime support of the test functions.
The problem of computing intersected von Neumann algebras is in general not
only difficult (since there are no known general computational techniques) but
also very unusual as compared to functional integral representation methods
related to Lagrangian quantization.

The problem becomes more amenable if one considers instead of operators their
formfactors i.e. their matrix elements between incoming ket and outgoing bra
state vectors. In the spirit of the old LSZ formalism one can then make an
Ansatz in form of a power series in $Z(\theta)$ and $Z^{\ast}(\theta)\equiv
Z(\theta-i\pi)$ (corresponding to the power series in the incoming free field
in LSZ theory). In a shorthand notation which combines both frequency parts we
may write
\begin{equation}
A=\sum\frac{1}{n!}\int_{C}...\int_{C}a_{n}(\theta_{1},...\theta_{n}%
):Z(\theta_{1})...Z(\theta_{n}):d\theta_{1}...d\theta_{n} \label{series}%
\end{equation}
where each integration path $C$ extends over the upper and lower part of the
rim of the $(0,-i\pi)$ strip in the complex $\theta$-plane. The
strip-analyticity of the coefficient functions $a_{n}$ expresses the
wedge-localization of $A\footnote{Compact localization leads to coefficient
functions which are meromorphic outside the open strip \cite{BFKZ}.}.$ It is
easy to see that these coefficients on the upper part of $C$ (the annihilation
part) are identical to the vacuum polarization form factors of $A$%
\begin{equation}
\left\langle \Omega\left|  A\right|  p_{n},..p_{1}\right\rangle ^{in}%
=a_{n}(\theta_{1},...\theta_{n}),\;\ \theta_{n}>\theta_{n-1}>...>\theta_{1}%
\end{equation}
whereas the crossing of some of the particles into the left hand bra state
(see the previous section) leads to the connected part of the formfactors
\begin{equation}
^{out}\left\langle p_{1},..p_{l}\left|  A\right|  p_{n},..p_{l+1}\right\rangle
_{conn}^{in}=a_{n}(\theta_{1}+i\pi,...\theta_{l}+i\pi,\theta_{l+1}%
,..\theta_{n})
\end{equation}
Hence the crossing property of formfactors is encoded into the notation of the
operator formalism (\ref{series}) in that there is only one analytic function
$a_n$ which describes the different possibilities of placing $\theta$ on the
upper or lower rim of $C.$ \ This is analogous to the
Glaser-Lehmann-Zimmermann expansion formulas \cite{GLZ} of the interacting
Heisenberg fields in terms of free fields in which the $n^{th}$ term is the
on-shell value of the Fourier transform of a retarded function which combines
the different formfactors for fixed $n$.

In terms of the formfactors the relative commutant (\ref{rel}) results from
restricting the series (\ref{series}) by requiring that the $A^{\prime}s$
commute with the generators of the shifted algebra $\mathcal{A}(W_{a})$%
\begin{equation}
\left[  A,U(a)\phi(f)U(a)^{\ast}\right]  =0 \label{com}%
\end{equation}

Thanks to the simplicity of the wedge generators $\phi(f),$ the $Z$ series of
the commutator can be computed in terms of the $a_{n}.$ The linearity of
$\phi(f)$ in the $Z^{\prime}s$ results in the $n^{th}$ term being a linear
combination of $a_{n-1}$ and $a_{n+1}.$ The denseness of $W$ $\ $localized
functions and the analyticity in the open strip finally lead to the
equivalence of the vanishing of this commutator with the famous ``kinematical
pole condition'', namely the $a_{n-1}$ function can be expressed as a residuum
of a pole in $a_{n+1}$%
\begin{equation}
Res_{\theta_{12}=i\pi}a_{n}(\theta_{1},\theta_{2},...\theta_{n})=2ia_{n-2}%
(\theta_{3},...\theta_{n})(1-S_{2n}...S_{23}),\,\theta_{12}=\theta_{1}%
-\theta_{2}%
\end{equation}
This relation was first postulated as one of the construction recipes by
Smirnov \cite{Smir}; it is the only relation between different components in
the absence of bound states. This together with the Payley-Wiener Schwartz
analytic characterization of the localization region and the crossing property
(which links the crossed formfactor to the analytic continuation between the
two rims of the $\theta$-strip $\mathbb{R}+i(0,\pi)$) characterizes the
\textit{space of formfactors} associated with the algebra $\mathcal{A}(D).$
Attempts to improve the localization by restricting the support of $f$ in the
$\mathcal{A}(W)$ generators $\phi(f)$ to a smaller region $suppf\subset
\mathcal{D}\subset W$ would fail; the generator continues to be
wedge-localized and by sharpening test function supports one can only enlarge
but not reduce the localization region.

The multiplicative structure\footnote{If the formfactors are matrixelements of
operators, they must also have a multiplicative structure which corresponds to
sums over infinitely many multi-particle states.} is outside of mathematical
control as long as one is unable to take care of the convergence of the
infinite sums; in this respect the situation is at first sight not better than
that of the old GLZ  expansion formulas \cite{GLZ} for interpolating
Heisenberg fields in terms of out/in free fields in which the coefficient
functions are on-shell restrictions of retarded correlation functions. The
linear space of formfactors can be parametrized in terms of a covariant basis
which corresponds to the formfactors of a basis of ``would be'' composite
fields. It turns out that the dependence on the individual composite field in
the Borchers class of relatively local fields can be encodes into a polynomial
factor \cite{BFKZ} (after splitting off a common factor which is the same for
all fields in the same class). This tells us that if we knew that those
operator subalgebras characterized by the vanishing of the relative commutant
(\ref{com}) are nontrivial, then the associated quantum field theory exists as
a algebraically nontrivial theory and we have a nonperturbative formalism to
compute formfactors of pointlike fields or of more general operators in
$\mathcal{A}(D).$ 

Since the formalism only involves formfactors but avoids correlation functions
of pointlike fields, it is free of ultraviolet problems (and a fortiori does
not require renormalization of infinities). Hence the world of factorizing
models is a candidate for the first explicit illustration of Pascual Jordan's
envisaged paradise of local quantum physics where one is able to walk without
classical crutches.

There has been extensive work on the calculation of formfactors of composite
fields. Similar to Wick polynomials there exists a basis of (composite) fields
in the same superselection sector. As mentioned, the formfactors of fields
from the same local equivalence class contain one factor which is common to
all of them (the so-called minimal formfactor \cite{BFKZ}); this factor is
associated with the ``core'' of the vacuum polarization cloud which is common
to all states created by operators from the same spacetime region and with the
same charge. It is this factor which carries the interaction; the remaining
polynomial factor is  in the exponential of the rapidities carries the
information about the different fields in the local equivalence class; this is
analogous to the different Wick polynomials of free fields.\footnote{This
factor is different for bounded operators $A\in\mathcal{A}(D)\,$where one
obtains a decrease for large momenta which may help in the control of the
convergence in (\ref{series}).}. The polynomial factors actually complicate
the calculation of correlation functions as convergent series over
formfactors. In fact apart from two-point functions in very special cases, the
program of controlling correlation functions of pointlike fields was \ without
much success, despite many attempts. The short distance aspects, which were
banned thanks to the on-shell nature of the bootstrap formfactor program,
enter through the back door in the form of convergence problems for the series
(\ref{series}).

In this context it is very interesting to note that recently Buchholz and
Lechner \cite{Bu-Le} proposed an elegant criterion for the nontriviality of
$\mathcal{A}(D)$ in terms of an operator algebraic property of the wedge
algebra $\mathcal{A}(W)$ which allows to bypass the problem of controlling
formfactor series altogether. They found that the ``nuclear modularity'' of
$\mathcal{A}(W)$ insures the nontriviality of the $\mathcal{A}(D)$
intersection and its standardness (the Reeh-Schlieder property) with respect
to the vacuum. Lechner tested this criterion in the case of the Ising field
theory \cite{Lech2}. There seems to be a well-founded hope that the already
impressive calculational results of the bootstrap-formfactor program for
factorizing models will be backed up by a structural argument of the existence
of their local algebras without having to control the convergence of infinite
sums over formfactors. Although the knowledge of wedge algebras already
determines the algebras associated with intersections uniquely, the
Buchholz-Lechner idea applies only to d=1+1 theories.

In the following two sections I will present ideas by which one hopes to
generalize the formfactor bootstrap approach.

\section{The hypothesis of a Masterfield}

For factorizable models, the crossing relation of the analytic coefficient
functions in the series representation (\ref{series}) is a consequence of the
algebraic properties of the $Z^{\prime}s.$ Since there are no Z-F operators
for models with non factorizing S-matrices, one must look for a more general
operational formulation of crossing. In order to obtain an idea in what
direction to look for, let us first recall the precise conceptual position of
factoring models within the general setting of massive models with a mass gap
(to which scattering theory applies).

As was mentioned in the second section, PFGs with generating properties for
wedge-localized algebras only exist for d=1+1 theories with S-matrices which
factorize into 2-particle contributions $S^{(2)}.$ This is a very peculiar
situation in which cluster separability does not distinguish between the two
contributions in $S^{(2)}=1+T^{(2)}$ since they carry the same energy-momentum
delta functions$.$

So the crucial question is how can one get an operational formulation of
crossing in formfactors\footnote{We always mean the connected part of the
formfactors which is what one gets by starting with the outgoing components of
the one field (or operator from a local algebra) state and crossing from
outgoing bras to incoming kets.} beyond such special situations? We already
dismissed the idea of interpreting crossing as KMS property in the same theory
as incorrect. The only alternative idea which maintains a KMS interpretation
of crossing, would consist in declaring simply the formfactors of an operator
$A$ to be correlation functions in a KMS state at the Hawking-Unruh
temperature $2\pi$ of (nonlocal) operators $R^{(A)}$ in rapidity momentum
space (the auxiliary $R^{\prime}s$ will be referred to as ``Rindler
operators'')
\begin{equation}
\left\langle \theta_{1},...\theta_{n}\left|  A\right|  \Omega\right\rangle
\overset{?}{=}\left\langle R^{(A)}(\theta_{1})...R^{(A)}(\theta_{n}%
)\right\rangle \label{master}%
\end{equation}
But this idea only works if we find special operators $A$ in the original
theory whose formfactors define a system of positive R-correlation functions,
since then the GNS reconstruction would lead to a global auxiliary operator
field theory. A necessary condition for such an interpretation is the validity
of the cluster separation property. It is known that this property holds also
in global operator algebras (i.e. algebras without a local net substructure)
as long as the operator algebra is a von Neumann factor in which case it is
related to the property of asymptotic abelieness \cite{Haag}\cite{Br-Ro}. In
many factorizable models one was able to identify such fields with rapidity
space clustering \cite{B-K}. We will formulate a requirement, which we call
the hypothesis of a ''masterfield''

\begin{definition}
A masterfield M(x) associated to a QFT is a distinguished scalar Boson field
within the Borchers class of locally equivalent fields whose connected
formfactors\textit{\ defines a thermal auxiliary ``Rindler QFT''\ at the
Hawking temperature }$\beta=2\pi$\textit{\ in terms of a nonlocal field
}$R(\theta,p_{\perp})$\textit{\ in the sense of the above formula
(\ref{master}) with }$A$\textit{\ being the masterfield M(x) at x=0.}
\end{definition}

The KMS relation in $\theta$ reads
\begin{align}
&  \left\langle R^{(M)}(\theta_{1},p_{1\perp})...R^{(M)}(\theta_{n-1}%
,p_{n-1\perp})R^{(M)}(\theta_{n},p_{n\perp})\right\rangle _{\beta=2\pi
}\label{R}\\
&  =\left\langle R^{(M)}(\theta_{n}-2\pi i,p_{n\perp})R^{(M)}(\theta
_{1},p_{1\perp})...R^{(M)}(\theta_{n-1},p_{n-1\perp})\right\rangle
_{\beta=2\pi}\nonumber\\
&  =(R^{(M)}(\theta_{n}-2\pi i,p_{n\perp})^{\ast}\Omega_{\beta=2\pi}%
,R^{(M)}(\theta_{1},p_{1\perp})...R^{(M)}(\theta_{n-1},p_{n-1\perp}%
)\Omega_{\beta=2\pi})\nonumber\\
&  =(JR^{(M)}(\theta_{n}-\pi i,p_{n\perp})\Omega_{\beta=2\pi},R^{(M)}%
(\theta_{1},p_{1\perp})...R^{(M)}(\theta_{n-1},p_{n-1\perp})\Omega_{\beta
=2\pi})\nonumber
\end{align}
where in the last two lines we used the more convenient state-vector notation
for the thermal expectation values and modular theory in order to convert
$\Delta^{\frac{1}{2}}$ into $J$. The identification of this expression with
the crossing property of the formfactor of $M(0)$%
\begin{align}
&  \left\langle 0\left|  M(0)\right|  p_{1},...p_{n}\right\rangle
=\left\langle -\bar{p}_{n}\left|  M(0)\right|  p_{1},...p_{n-1}\right\rangle
\\
&  =(JR^{(M)}(\theta_{n}-\pi i,p_{n\perp})\Omega_{\beta=2\pi},R^{(M)}%
(\theta_{1},p_{1\perp})...R^{(M)}(\theta_{n-1},p_{n-1\perp})\Omega_{\beta
=2\pi})\nonumber
\end{align}
requires the action of the auxiliary $J$ as $JR^{(M)}(\theta_{n}-\pi
i,p_{n\perp})\Omega_{\beta=2\pi}=R^{(M)}(\theta_{n}-2\pi i,-p_{n\perp})^{\ast
}\Omega_{\beta=2\pi}.$ In d=1+1 the interpretation of crossing in terms of KMS
of an auxiliary theory simplifies, since there is no transverse momentum
$p_{\perp}$.

It is important to notice that the auxiliary field theory associated with the
formfactors of the master field is not subject to the restriction of
wedge-localized PFGs which led to factorizable models. In fact being a global
(i.e. without a local net structure) KMS theory, the concept of particles and
in particular the concept of PFG becomes meaningless.

Let us first look at the rather trivial illustration of a free master field
namely
\begin{align}
&  M(x)\equiv:e^{\gamma A(x)}:,\,\,A(x)=free\,field\\
&  \left(  \Omega\left|  M(0)\right|  p(\theta_{1},p_{1,\perp})...p(\theta
_{n},p_{n,\perp})\right)  =e^{c\gamma}\nonumber
\end{align}
where the positive constant $c$ is related to the vacuum-one particle
normalization of $A.$ Clearly among all composites of the free field which
lead to $\theta$-independent connected formfactors, the only case with the
correct combinatorics complying with clustering is the above exponential
field. The auxiliary algebra of $R^{(M)}$ is the trivial abelian algebra which
permits states for every KMS temperature. The free field is also the only
model in which the formfactors of the masterfield define an abelian auxiliary
theory; a nontrivial S-matrix prevents abelienness.

The masterfield hypothesis remains nontrivial even in the setting of
factorizable models. In the following we use two quite different models to
illustrate its working. We first recall some formalism of KMS states on free fields.

For bosonic quasifree KMS states at the KMS temperature $\beta$ one obtains
\begin{align}
&  \left\langle c(q)c(q^{\prime})\right\rangle _{\beta}=e^{\beta q}\left\{
\left\langle c(q)c(q^{\prime})\right\rangle _{\beta}-\left[  c(q),c(q^{\prime
})\right]  \right\} \\
&  \curvearrowright\left\langle c(q)c(q^{\prime})\right\rangle _{\beta}%
=\frac{e^{\beta q}}{e^{\beta q}-1}\varepsilon(q)\delta(q+q^{\prime})\nonumber
\end{align}

For the first illustration we take the Sinh-Gordon theory. The field which
leads to formfactors which have the cluster factorization property in the
rapidity variable is again an exponential $M(x)=e^{\varphi}$ operator in terms
of the basic Sinh-Gordon field $\varphi$ \cite{B-K}. They are known to have
the following structure
\begin{align}
&  \left(  \Omega\left|  M(0)\right|  p(\theta_{1})...p(\theta_{n})\right)
=K_{n}(\underline{\mathbf{\theta}})\prod_{i<j\leq n}F(\theta_{ij})\\
&  K_{n}(\underline{\mathbf{\theta}})=\sum_{l_{1}=0,1}...\sum_{l_{n}%
=0,1}(-1)^{\sum l_{i}}\prod_{i<j}(1+(l_{i}-l_{j})\frac{i\sin\pi\nu}%
{\sinh\theta_{ij}})\prod_{k}Ce^{i\pi\frac{\gamma}{\beta}(-1)^{l_{k}}}\nonumber
\end{align}
where the coupling strength $\beta$ and $\nu$ are related by $\frac{1}{\nu
}=\frac{8\pi}{\beta^{2}}.$ The product factor involves the 2-particle
formfactor $F$ and has the combinatorics of an exponential which is bilinear
in $c(q)$ free Boson operators. This suggests to start from the complex
exponential
\begin{align}
C(\theta)  &  =e^{ia(\theta)}\\
a(\theta)  &  =\int dqw(q)c(q)e^{iq(\theta-i\frac{\pi}{2})}\nonumber
\end{align}
and look for a Rindler operator $R\footnote{We use the letter $Z$ for the
particle physics representation of the Zamolodchikov algebra (the Minkowski
spacetime operators which are related to the PFG wedge generators) whereas $R$
is used for the thermal Rindler representation.}$ as a Hermitian combination
of the form
\begin{align}
&  R(\theta)=N\left\{  e^{i\gamma}C(\theta-i\frac{\pi}{2})+h.a.\right\}
\label{C}\\
&  C(\theta)C(\theta^{\prime})=S^{(2)}(\theta-\theta^{\prime})C(\theta
^{\prime})C(\theta),\text{\thinspace}S^{(2)}(\theta)=exp\int_{0}^{\infty
}dqf(q)sinhq\frac{\theta}{i\pi}\nonumber\\
&  \left\langle C(\theta-i\frac{\pi}{2})C(\theta^{\prime}-i\frac{\pi}%
{2})\right\rangle _{2\pi}=exp\int_{0}^{\infty}dqf(q)\frac{1-chq(\pi
+i(\theta-\theta^{\prime}))}{2sh\pi q}\nonumber\\
&  \equiv F_{min}(\theta-\theta^{\prime})\nonumber
\end{align}

The function $F_{min}(\theta)$ is the so-called minimal 2-particle formfactor
of the model i.e. the unique function which obeys $F(\theta)=S^{(2)}%
(\theta)F(-\theta)$ and is holomorphic in the strip. For the present model
without bound states it agrees with $F.$ In the last line in (\ref{C}) we used
the fact that the KMS state at the inverse temperature $2\pi$ fixes the
quasi-free state on the Rindler creation/annihilation operator algebra which
in turn determines the thermal expectations of the $C$-operators.

The Sinh-Gordon S-matrix
\begin{equation}
S_{sh}(\theta)=\frac{th\frac{1}{2}(\theta-i\kappa)}{th\frac{1}{2}%
(\theta+i\kappa)},\,\,\kappa=\frac{\pi\beta^{2}}{8\pi+\beta^{2}}=\pi B\leq\pi
\end{equation}
fixes the quasifree commutation relation of the Rindler operators $R(\theta)$
with
\begin{equation}
f(q)=\frac{2sh\frac{\kappa q}{2}sh\frac{\kappa q8\pi}{2\beta^{2}}}%
{qch\frac{\pi q}{2}}=\frac{2sh\frac{q\pi}{2}Bsh\frac{q\pi}{2}(2-B)}%
{qch\frac{\pi q}{2}}%
\end{equation}

The n-point function
\begin{equation}
\left\langle C(\theta_{1}-i\frac{\pi}{2})....C(\theta_{n}-i\frac{\pi}%
{2})\right\rangle _{2\pi}\sim\prod_{i<k}F_{min}(\theta_{ik})
\end{equation}
fulfills the commutation relation of the $R$-algebra (which is identical to
that of the $C$-algebra as well as the KMS condition. Our interest lies in the
Hermitian field operator $R.$ For convenience we have adjusted our notation to
the resulting combinatorics for the thermal $Z$-expectation which are sums of
terms with different $C_{l}(\theta):=C(\theta-il\frac{\pi}{2}),$ $l=\pm$
\begin{align}
\left\langle C_{l_{n}}...C_{l_{1}}\right\rangle _{2\pi}  &  =\prod
_{i<k}F_{min}(\theta_{ik})\left\{  1-(l_{k}-l_{i})\frac{isin\kappa}%
{sh\theta_{ki}}\right\}  ,\,\,l_{i}=\pm1\\
\left\langle R(\theta_{n})...R(\theta_{1})\right\rangle _{\beta=2\pi}  &
\sim\sum\left\langle C_{l_{n}}...C_{l_{1}}\right\rangle _{2\pi}expi\pi
\alpha(l_{1}+l_{2}+...l_{n})\nonumber
\end{align}
where $\alpha$ depends on the numerical pre-factors.

With the present construction of an auxiliary global Rindler QFT for the
formfactors of the masterfield we have reproduced a curious observation by
Lukyanov \cite{Lukyanov} which is known in the literature on factorizing
models as ``free field representations'' (for a recent account see also
\cite{Lash}). The difference to Lukyanov is in the underlying concepts and not
in the actual computation. The thermal state turned out to be a Rindler-Unruh
KMS state at a fixed Hawking temperature rather than a tracial Gibbs states in
a heat bath setting. Unique KMS states on operator algebras lead to von
Neumann factors which in turn fulfill weak cluster property \cite{Br-Ro} and
it was the cyclicity of crossing together with the somewhat mysterious cluster
properties in the rapidity variables \cite{B-K} which suggested this operator
KMS interpretation of the crossing property for the formfactors of a masterfield.

Since fields whose formfactors cluster have been found in many similar
factorizing models of Toda type \footnote{In \cite{B-K3} it was shown that
distinguished fields with clustering formfactors exist for all $A_{n-1}$
affine Toda field theories of which the Sinh-Gordon is the lowest member.},
one would expect that the idea of an auxiliary Rindler theory in momentum
space works in all of them. Moreover it would be tempting to conjecture that
the simplifying feature of factorizing models consists in the auxiliary
formfactor theory being bilinear exponential in $c^{\#}(q)$
creation/annihilation operators. This conjecture draws also support from a
recent observation by Babujian and Karowski who observed that a suitably
generalized form of clustering for also holds in statistics changing $Z_{n}%
$-models \cite{B-K3} of which the lowest one is the Ising field theory. In
that case a combination of disorder/order field formfactors leads to
clustering \cite{Cherno}.

In the following we briefly show that the masterfield idea also works in the
Ising model; in that case the relevant state is a ``twisted'' KMS state at the
temperature $\beta=\pi$. The twisting consists in changing the KMS formula by
a -sign.
\begin{align}
\left\langle c(q)c(q^{\prime})\right\rangle _{\beta}  &  =-e^{\beta q}\left\{
\left\langle c(q)c(q^{\prime})\right\rangle _{\beta}-\left[  c(q),c(q^{\prime
})\right]  \right\} \\
\left\langle c(q)c(q^{\prime})\right\rangle _{\beta}  &  =\frac{e^{\beta q}%
}{1+e^{\beta q}}\varepsilon(q)\delta(q+q^{\prime})\nonumber\\
\left\langle c(q)c(q^{\prime})\right\rangle _{\pi}  &  =\frac{e^{\frac{\pi}%
{2}q}}{2cosh\frac{\pi}{2}q}\varepsilon(q)\delta(q+q^{\prime})\nonumber
\end{align}
where the third line contains the KMS two-point function at $\beta=\pi$ which
we are going to use together with the following definition of $a(\theta)$%
\begin{align}
a(\theta)  &  =\int c(q)dq\\
\left\langle a(\theta)a(\theta^{\prime})\right\rangle _{\beta=\pi}  &
=\int_{0}^{\infty}\frac{\sinh q(i\theta-i\theta^{\prime}+\frac{\pi}{2})}%
{\cosh\frac{\pi}{2}q}\frac{dq}{q}=\ln i\tanh\frac{\theta-\theta^{\prime}}%
{2}\nonumber
\end{align}
which finally leads to the well-known disorder/order Ising formfactors which
is given by a combinatorial expression in the two-particle formfactor of the
disorder operator (which correspond to an even number of particles)
\begin{equation}
\left\langle e^{a(\theta)}e^{a(\theta^{\prime})}\right\rangle _{\pi}%
=\tanh\frac{\theta-\theta^{\prime}}{2}%
\end{equation}

As the Sinh-Gordon model is the simplest representative of the class of
$A_{n-1}$ affine Toda models \cite{B-K3}, the Ising field theory is the first
in the family of $Z_{n}$ models. These models are more difficult as a
consequence of their preferred $Z_{n}$ braid group statistics and a candidate
for a masterfield is not immediately visible. The suggestion from the Ising
case would be that a suitable combination of all disorder/order operators
would be a candidate for a field which fulfills some generalized clustering
(i.e. adjusted to the exotic statistics).

The important point underlying the idea of a masterfield is that there exists
an analytic correlation function (\ref{R}) whose different boundary values in
momentum rapidity space (\ref{R}) correspond to different operator ordering.
For factorizing $S_{scat}$ matrices the close relation between transpositions
and actions of $S^{(2)}$ suggested how to relate the different orderings to
on-shell operator data. For general $S_{scat}$-matrices we could the opposite
$\theta$-order with the action of $S_{scat}$%
\begin{equation}
\left|  n,n-1,...2,1\right\rangle =S_{scat}\left|  1,2,...n-1,n\right\rangle
\end{equation}
but it is not obvious what kind of operator relation one should use for other
orderings. Perhaps the cluster property leads to further restrictions which
together with the KMS property permit to determine the auxiliary R-theory. In
any case it seems to me that an operator interpretation of the different
rapidity orderings in formfactors and the KMS property is an indispensable
part of a deeper operator understanding of crossing and a (perturbative)
on-shell construction.

The on-shell bootstrap-formfactor idea is not the only possibility to avoid
short-distance problems resulting from the use of field coordinatizations and
their singular correlations. Another less speculative but by no means simpler
idea will be presented in the next section.

\section{Lightfront Holography as a constructive tool?}

In the previous sections we have used modular theory together with on-shell
concepts in order to analyze wedge algebras in the presence of interactions.
In this section I will present a recent proposal which also uses modular
localization ideas in order to simplify the problem of classifying and
constructing QFTs. But instead of particle concepts, as e.g. PFGs for wedge
algebras and formfactors, it is based on the good understanding of chiral
theories which are related to the actual theory by a process of ``algebraic
lightfront holography'' (ALH).

Of course no non-perturbative approach to higher dimensional interacting QFT
can achieve miracles; simplification just means the partition of a complex
dynamical problem into a sequence of less complicated single steps. Perhaps
the following comparison with the canonical formalism sheds additional light
on this point. This ETC formalism tries to classify and construct QFTs by
assuming the validity of canonical equal time commutation relations (ETCR).
The shortcomings of that approach are well-known. Apart from the fact that in
higher dimensional relativistic QFT the ETC structure is inconsistent with the
presence of interactions, ETCR are not useful as a starting point for a rough
intrinsic distinction between different (universality) classes of interactions
since ETCR are totally universal.

Lightfront holography tries to address this imbalance by replacing the ETCR by
the much richer structure of chiral theories on the lightfront. Starting from
a richer ``kinematical'' setting than ETCR, one may hope for a more accessible
``dynamical'' side. The holographic projection may map different ambient
theories to the same chiral image, but similar to the better known scale
invariant short distance universality classes, the holographic universality
classes allow for more realizations than the unique ETCR structure. However in
contradistinction from scaling limits, holographic projections live in the
same Hilbert space as the ambient theory; in fact they just organize the
spacetime aspects of a shared algebraic structure in a radically different way.

Let us briefly recall the salient points of ALH\footnote{We add this prefix
``algebraic'' in order to distinguish the present notion of holography from
the gravitational holography of t'Hooft \cite{Hooft}. More on similarities and
differences between the two can be found in the concluding remarks,}.

ALH may be viewed as a kind of conceptually and mathematically updated
``lightcone quantization'' (or ``$p\rightarrow\infty$ frame'' description).
Whereas the latter approaches never faced up to the question of how the new
fields produced by the lightfront quantization prescriptions are related to
the original local fields i.e. in which sense the new description addresses
the original problems posed by the ambient theory, the ALH is conceptually
precise and mathematically rigorous on this points. It turns out that the idea
of restricting fields to the lightfront is limited to free fields and certain
superrenormalizable interacting models with finite wave function
renormalization (which only can be realized in d=1+1). Theories with
interaction-caused vacuum polarization which leads to Kallen-Lehmann spectral
functions with diverging wave function renormalization factors do not permit
lightfront restrictions for the same reason as they do not have equal time
restrictions; e.g. for scalar fields on has\footnote{It is important to
realize that the LF restriction is not a pointwise procedure. The best
understanding is achieved within the setting of modular localization (see
below).}
\begin{align}
\left\langle A(x)A(y)\right\rangle  &  =\int_{0}^{\infty}\rho(\kappa
^{2})i\Delta^{(+)}(x-y,\kappa^{2})d\kappa^{2}\\
\left\langle A(x)A(y)\right\rangle |_{LF} &  \sim\int_{0}^{\infty}\rho
(\kappa^{2})d\kappa^{2}\delta(x_{\perp}-y_{\perp})\int_{0}^{\infty}\frac
{dk}{k}e^{-ik(x_{+}-y_{+})}\nonumber
\end{align}
where in passing to the second line we used the rule (\ref{lim}) which
replaces the ambient two-point function of mass $\kappa$ by its zero mass
lightfront restriction in the sense of the second section. As explained there,
the infrared-divergence in the longitudinal factor is spurious if one views
the lightfront localization in the setting of modular wedge localization. On
the other hand the obstruction resulting from the large $\kappa$ divergence of
the K-L spectral function (short distance regime of interaction-caused vacuum
polarization) is shared with that in ETCR i.e. in both cases the process of
restriction is meaningless. 

However whereas equal time restricted interacting fields in d=1+3 simply do
not exist, there is no such limitation on the short distance properties of
generalized chiral conformal fields which turn out to generate the ALH. What
breaks down is only the idea that these lightfront generating fields can be
gotten simply by restricting the fields of the ambient theory, as was the case
in the example of free fields in the second section.

It turns out that in algebraic lightfront holography the connection between
the ambient theory and its holographic projection requires the use of modular
theory. Although the ambient theory may well be given in terms of pointlike
fields and the ALH may also allow a pointlike description (see \ref{gen}),
there is no direct relation between these fields. This also sheds light on the
old difficulties with lightcone quantization which posed an obstacle to
generations of physicists; even in the interaction-free case when the
restriction works, the ALH net of algebras is nonlocal relative to the ambient
algebra and hence the recovery of the ambient from the LF operators involves
nonlocal steps. Whereas lightcone quantization was not able to address those
subtle problems, ALH solves them.

The intuitive physical basis of this algebraic approach is a limiting form of
the causal closure property. Let $\mathcal{O}$ be a spacetime region and
$\mathcal{O}^{\prime\prime}$ its causal closure (the causal disjoint taken
subsequently taken twice) then the causal closure property is the following
equality
\begin{equation}
\mathcal{A}(\mathcal{O})=\mathcal{A}(\mathcal{O}^{\prime\prime})
\end{equation}
In the case of free fields this abstract algebraic property is inherited via
quantization from the Cauchy propagation in the classical setting of
hyperbolic differential equations. The lightfront is a limiting case
(characteristic surface) of a Cauchy surface. Each lightray which passes
through $\mathcal{O}$ either must have passed or will pass through
$O^{\prime\prime}.$ For the case of a $x^{0}-x^{3}$ wedge $W$ and its
$x^{0}-x^{3}=0$ (upper) causal lightfront boundary $LFB(W)$ (which is half of
a lightfront) the relation
\begin{equation}
\mathcal{A}(LFB(W))=\mathcal{A}(W)\label{LFB}%
\end{equation}
is a limiting situation of the causal shadow property; a lightlike signal
which goes through this boundary must have passed through the wedge (or in the
terminology of causality, the wedge is the backward causal completion of its
lightfront boundary). Classical data on the lightfront define a characteristic
initial value problem and the smallest region which casts an ambient causal
shadow is half the lightfront as in (\ref{LFB}); for any transversely not
two-sided infinite extended subregion, as well as for any region which is
bounded in the lightray direction, the causal completion is trivial i.e.
$\mathcal{O=O}^{\prime\prime}.$ This unusual behavior of the lightfront is
related the fact that as a manifold with its metric structure inherited from
the ambient Minkowski spacetime it is not even locally hyperbolic.

Some of the symmetries which the lightfront inherits from the ambient
Poincar\'{e} group are obvious. It is clear that the lightlike translation
together with the two transverse translation and the transverse rotation are
leaving the lightfront invariant and that the longitudinal Lorentz boost,
which leaves the wedge invariant, acts as a dilatation on the lightray in the
lightfront. There are two additional invariance transformations of the
lightfront which are less obvious. Their significance in the ambient space is
that of the two ``translations'' in the 3-parametric Wigner little group
$E(2)$ (a Euclidean subgroup of the 6-parametric Lorentz group) which leave
the lightray in the lightfront invariant. Projected into the lightfront these
``translations'' look like transverse Galilei transformations in the various
$(x_{\perp})_{i}-x_{+}$ planes.

The resulting 7-parametric symmetry group of the lightfront is used to
construct the modular localization structure of the ALH. For the longitudinal
localization in the lightray direction the construction is based on the
inclusion \cite{Wies}\cite{Sch-Wies}
\begin{equation}
\mathcal{A}(W)\supset\mathcal{A}(W_{e_{+}})\equiv AdU(e_{+})\mathcal{A}(W)
\end{equation}
where $\mathcal{A}(W_{e_{+}})$ is the image of $\mathcal{A}(W)$ under a
translation $e_{+}$ along the lightray. This inclusion is known to be
``half-sided modular'' (hsm) i.e. the modular group of the larger algebra
$\Delta_{W}^{it}$ compresses the smaller one for $t<0$ (+ half-sided modular)
\begin{equation}
Ad\Delta_{W}^{it}\mathcal{A}(W_{e_{+}})\subset\mathcal{A}(W_{e_{+}}),\,\ t<0
\end{equation}
It is well-known \cite{G-L-W} that such inclusions lead to Moebius covariant
chiral nets precisely if they are ``standard'' i.e. if
\begin{equation}
\mathcal{A}(W_{e_{+}})^{\prime}\cap\mathcal{A}(W)\Omega
\,\,is\,\,dense\text{\thinspace}in\text{ }H
\end{equation}
The lightlike inclusion is the limit of spacelike inclusions which in
compactly localizable theories are evidently standard (but not hsm). This
property is known to hold in the absence of interactions where it can be
traced back to the spatial standardness of the respective subspaces of the
Wigner representation space \cite{GLRV}. For factorizing models in d=1+1 this
algebraic requirement is the prerequisite for the existence of pointlike
fields in the bootstrap formfactor approach. The fact that the short distance
behavior of these fields admit arbitrary high inverse powers suggests that
this standardness assumption (unlike the lightcone quantization and the above
lightfront restriction method) is not affected by short distance properties.
Since our aim is the classification and construction of models, the range of
validity of our method is at the end decided by its future success.

The interpretation of the chiral net obtained from the hsm inclusion for d%
%TCIMACRO{\TEXTsymbol{>}}%
%BeginExpansion
$>$%
%EndExpansion
1+1 is that of a system of algebras associated with transverse ``slices''
(stripes in case of d=1+2) i.e. regions of finite longitudinal and two-sided
infinite transverse extension. Note that the conformal rotation (or the proper
conformal transformation), which requires the one-point compactification of
the longitudinal coordinate, does not arise from the holographic projection of
the Poincar\'{e} transformations, but rather results from the
symmetry-improving aspect of lightfront holography \cite{Verch}.

In order to obtain the complete local resolution on the lightfront we still
have to find a mechanism which generates a transverse localization structure.
This is done with the help of ``modular intersections''. For this purpose we
now use the aforementioned two ``translations'' in Wigner's little group
$E(2).$ These transformations tilt the wedge $W$ in such a way that its upper
boundary remains inside the lightfront. The thickness of the slice in the
lightray direction is maintained whereas the transverse directions are tilted
in the sense of Galilei group actions. It is easy to see that the intersection
of the algebras localized in the original slice with those of the tilted slice
defines an algebra localized in a finite region. The net structure of the
lightfront algebra is defined in terms of this intersected slice algebras. As
modular inclusions of wedges are inexorably linked to dilation-translation
symmetries, modular intersections of wedges are related to Wigner's little
group $E(2)$ \cite{Sch-Wies2}$\cite{hol}\cite{B-M-S}.$ For more on the
operator algebraic aspects of modular intersections we refer to the literature
\cite{Wies2}.

. The holographic projection method confirms that the vacuum polarization
properties, which for free fields can be explicitly derived by the lightfront
restriction method, continue to hold in the presence of interactions. The most
surprising aspect is certainly the total absence of transverse vacuum
polarization which is a consequence of the following theorem on tensor
factorization \cite{Borchers}

\begin{theorem}
A von Neumann subalgebra $\mathcal{A}$ of $B(H)$ which admits a two-sided
lightlike translation with positive generator is of type I , i.e. it tensor
factorizes as $B(H)=\mathcal{A}\otimes\mathcal{A}^{\prime}$ associated with
$H=H_{\mathcal{A}}\otimes H_{\mathcal{A}^{\prime}}$ and the factorization of
the vacuum vector $\Omega=\Omega_{\mathcal{A}}\otimes\Omega_{\mathcal{A}%
^{\prime}}$
\end{theorem}

The transverse tensor factorization is corroborated by the application of the
Takesaki theorem \cite{Takesaki} which fits nicely into our modular based
approach since it relates the existence of preservation of subalgebras under
the action of the modular group of the ambient algebra to the existence of
conditional expectations.

\begin{theorem}
The modular group of an operator algebra in standard position ($\mathcal{B}%
,\Omega)$ leaves a subalgebra $\mathcal{A}\subset\mathcal{B}$ invariant if and
only if there exists a $\Omega$-preserving conditional expectation
$E:\mathcal{B}\rightarrow\mathcal{A}$. In that case the state $\omega
(\cdot)=(\Omega,\cdot\Omega)$ is a factor state on $\mathcal{A}\times
\mathcal{C}$ with$\ \mathcal{C}\equiv\mathcal{A}^{\prime}\cap\mathcal{B}$
which leads a tensor factorization $H_{\mathcal{B}}=H_{\mathcal{A}}\otimes
H_{\mathcal{C}}$ where the Hilbert spaces are cyclically generated from
$\Omega$ by the application of the respective algebras.
\end{theorem}

In the adaptation of this theorem to our problem we only have to set
$\mathcal{B}=\mathcal{A}(LFB(W)),$ $\mathcal{A}=\mathcal{A}(x_{\perp}\in
Q,x_{+}>0),$ where $Q$ is a compact region in the transverse coordinates$.$

These theorems clearly show that the holographic lightfront projection has a
transverse quantum mechanical structure since tensor factorization upon
subdivisions of spatial regions and factorization of the vacuum vector are the
characteristic features of QM. This unexpected property of encountering
quantum mechanical structures in relativistic QFT without having done any
nonrelativistic approximation is a characteristic property of ALH. It is
certainly related to the fact that the LF is not hyperbolic.

In addition to those symmetries inherited from the ambient theory there are
new symmetries as the result of the ``symmetry-improving'' lightfront
projection \cite{Verch}. One of them is the vacuum-preserving conformal
rotation (see later section for more comments).

It is interesting and useful to ask what kind of generating pointlike fields
$\psi$ could describe a holographic projection. The possibilities are severely
limited by the transverse tensor factorization and the longitudinal chiral
structure; they essentially amount to the following commutations structure
(can be easily extended to include fermionic operators)
\begin{align}
& \left[  \psi_{i}(x_{\perp},x_{+}),\psi_{j}(x_{\perp}^{\prime},x_{+}^{\prime
})\right]  =\label{gen}\\
& =\delta(x_{\perp}-x_{\perp}^{\prime})\left\{  \delta^{(n_{ij})}(x_{+}%
-x_{+}^{\prime})+\sum_{k}\delta^{(n_{ijk})}(x_{+}-x_{+}^{\prime})\psi
_{k}(x_{\perp},x_{+})\right\}  \nonumber
\end{align}
where the common $\delta$-function in the transverse direction takes care of
the quantum mechanical property and the longitudinal structure parallels that
known from the Lie-field structure of chiral observable algebras i.e. the
$\psi_{i}$ constitute a (finite or infinite) Lie-field basis and the sum
extends over finitely many terms. As in the pure chiral case of W-algebras the
number of the derivatives in the longitudinal $\delta$-functions is controlled
by the balance of scale dimensions on both sides of the equation.

The operators obtained by smearing with test functions $f(x_{\perp},x_{+})$
clearly produces the transverse quantum mechanical factorization as a result
of the presence of just one $\delta$-function without derivatives. Observables
with nonoverlapping transverse extension factorize according to%

\begin{equation}
\left\langle AA^{\prime}\right\rangle =\left\langle A\right\rangle
\left\langle A^{\prime}\right\rangle
\end{equation}

I believe that the existence of generating fields (\ref{gen}) for ALH can be
similarly argued as in \cite{Joerss} where generating fields for ordinary
chiral nets of algebras (without transverse extension) were constructed.

It is well-known that the wedge localization, and hence also the localization
on its causal boundary $LFB(W),$ causes a thermal behavior; in more specific
terms, the restriction of the vacuum to those localized algebras is
indistinguishable from a thermal KMS state at a fixed temperature (the Unruh
analog of the Hawking temperature) whose Hamiltonian is the generator of the
W-affiliated Lorentz boost. The temperature for the thermal aspects caused by
the quantum field theoretic vacuum polarization aspects at the boundary of
localization regions is related to the geometry of these regions; this is in
marked contrast to the standard heat bath thermality which leads to freely
variable temperatures and also exists in the classical setting. The absence of
vacuum polarization in the transverse direction suggests that the
localization-caused thermality leads to an entropy density i.e. an entropy per
unit transverse volume which has the dimension of an area \cite{hol}. This is
in marked contrast to the volume density of heat bath thermality and may well
turn out to be the QFT prerequisite for the Bekenstein area law in the
quasiclassical treatment of black holes.

In a constructive use of these ideas one would start with a classification of
QFT on the lightfront in terms of extended chiral theories and aim at the
reconstruction of the ambient theory as a kind of inverse ALH. The action of
the 7-parametric invariance subgroup on the lightfront algebra is part of the
ALH data. Their could be a restriction on the AHL data from the requirement
that the three remaining transformation which together with the LF invariance
group generate the ambient Poincar\'{e} symmetry act in a consistent way. In
analogy with the many Hamiltonians one Certainly one has to expect many ways
of Having arrived at the family of wedge algebras in terms of the ALH extended
chiral algebra the remaining construction of the ambient algebraic net is then
uniquely determined in terms of intersections.

Further inside can be gained by comparing the particle-based modular approach
to factorizing models with ALH. The representation of the generators
(\ref{PFG}) in terms of on-shell Z-F creation/annihilation operators
simplifies the calculation of the lightray limit $x_{-}=0.$ The method of
lightfront restriction works exactly as in the case of d=1+1 free fields
(\ref{lim}) except that corresponding formulas in terms of the Z-F operators
only serve as generators of half-line algebras. The algebras of finite
intervals have to be calculated as relative commutants by the modular
inclusion formalism; the resulting infinite series in the Z-F operators are
completely analogous to (\ref{series}) in section 4. In terms of pointlike
generating fields one has ($p_{-}(\theta)=me^{\theta}$)%
\begin{align}
A_{LF}(x_{+})  & =\sum\frac{1}{n!}\int_{C}...\int_{C}e^{ix_{+}(p_{-}%
(\theta_{1})+...p_{-}(\theta_{n}))}a_{n}(\theta_{1},...\theta_{n})\cdot\\
& :Z(\theta_{1})...Z(\theta_{n}):d\theta_{1}...d\theta_{n}\nonumber
\end{align}
The corresponding ambient massive pointlike localized fields are of the form
($p_{+}(\theta)=me^{-\theta}$)%
\begin{align}
A(x) &  =\sum\frac{1}{n!}\int_{C}..\int_{C}e^{ix_{+}(p_{-}(\theta
_{1})+..+p_{-}(\theta_{n}))+ix_{-}(p_{+}(\theta_{1}+..+p_{+}(\theta_{n})}%
a_{n}(\theta_{1},...\theta_{n})\\
&  :Z(\theta_{1})...Z(\theta_{n}):d\theta_{1}..d\theta_{n}\nonumber\\
p_{+}p_{-} &  =m^{2}\nonumber
\end{align}
The additional exponents involving the total $P_{+}$ momentum can be thought
of originating from a nonlocal ``Hamiltonian'' propagation law of the form
e$^{iP_{+}x_{-}}$

Apparently those chiral theories which arise as ALH projections\footnote{The
reader should note that the relation between the holographic chiral projection
and the ambient factorizing model is exact, whereas Zamolodchikov's working
hypothesis is based on a construction of factorizing models from their chiral
scaling limits by specific perturbations. Nevertheless there may be connection
between holographic and scaling universality classes.} from factorizing models
(and hence have PFGs in terms of Z-F variables) have a LF restriction which in
terms of these variables is similar to that for free fields. In particular the
covariance of the $Z^{\prime}s$ renders the extension into the ambient $x_{-}$
direction unique. As mentioned before we do not expect such a uniqueness of
the inverse lightfront holography beyond factorizing models, in particular for
higher dimensional theories.

The calculation of the intersection spaces associated with intervals on the
lightray is entirely analogous to that of the double cone intersections, in
both cases one obtains infinite series (\ref{series}) which applied to the
vacuum lead to rich vacuum polarization clouds. This interplay between a
massive 2-dimensional and chiral models is a new aspect of QFT since it does
not depend on any approximation or scaling limit and is therefore somewhat
surprising. It shows that at least certain chiral theories admit novel
descriptions in terms of a 2-dimensional particle basis. Whereas the
dilation-translation subgroup of the Moebius group leaves the vacuum as well
as the holographic images of the massive one-particle states invariant, the
Moebius rotation leaves the vacuum invariant but adds vacuum polarization
clouds to the alias one-particle states. More investigations on this
interesting point are required.

As a result of insufficient knowledge about higher dimensional models, there
is presently no model illustration of the ALH in the presence of transverse directions.

\section{\bigskip Concluding remarks}

In these notes we have been exploring nonperturbative ideas to access QFT
without using the classical ``crutches'' inherent in Lagrangian quantization
and without being subject to the severe short distance restrictions of the
related canonical commutation relations or those of the euclidean functional
integral representations\footnote{Functional integral representations suffer
the same limitations (for the same mathematical reasons) for interacting QFT
as the previously explained limitations of ETCR.} to make sense outside QM. 

A common aspect is the important role which modular localization plays in
these attempts. Without interactions, modular particle and field localizations
are functorially related as expressed by the ``commuting square''
(\ref{square}), but as a result of interaction-induced vacuum polarization the
particle localization is lost apart from the existence of wedge-localized PFGs
i.e. wedge-localized operators which applied to the vacuum create one-particle
states free of vacuum polarization admixtures. If one in addition requires
these operators to have reasonable domain properties with respect to
translations (tempered PFGs), only the d=1+1 factorizing models remain. In the
latter case it is also possible to formulate a quantum field theory of the
system of formfactors of a distinguished field called the masterfield. Whether
this masterfield idea has a higher dimensional generalization remains a matter
of speculation.

An interesting link between the old S-matrix bootstrap program and the
formfactor approach to QFT is the uniqueness of the inverse scattering problem
in QFT. Although it says nothing about the existence of a QFT, it at least
shows that if formfactors fulfill the crossing property, there is only one
local off-shell extrapolation i.e. only one local net with a given
$S_{scat}.\,\,$This is interesting in view of the historical relations of
string theory to Veneziano's dual model in which crossing property was
implemented with infinitely many particle states. Although this is quite
distinct from how crossing is expected to be achieved in QFT where both the
particle poles and the cuts from the scattering continuum enter the crossing
relation (as can be exemplified by the S-matrices of factorizing models), the
idea that the string prescriptions may turn out to be a just a differently
formulated local quantum physics was never totally ruled out\footnote{Actually
the canonical second quantization of the classical Nambu-Goto string leads to
pointlike local objects \cite{Martinec}\cite{Dimock}.}, despite many
conflicting opinions. 

One would feel more confidendent about this point, if crossing would have
continued to play the same pivotal role in string theory as it did in the
(genus zero) formulation of the dual model. But a glance at contemporary
string theory indicates that it dropped out of sight; it is not even clear
whether it holds at all. In this conceptually somewhat opaque situation it is
interesting to note that very recently the local quantum physics
interpretation of bosonic string field theory received some support from one
of the protagonists of string theory \cite{Er-Gr} by indicating the possible
construction of a (presumably infinite) set of local fields which interpolate
the string field theory S-matrix. In the spirit of ``intrinsicness'' set
forward in the present work, one might add the remark that by investigating
the crossing property associated with such an S-matrix, the uniqueness of the
inverse scattering problem based on crossing secures the uniqueness of the
associated local quantum physics in a way which does not depend on the art
(and luck) of finding local interpolating fields.

An alternative idea would be that the relevant objects of string field theory
are really semiinfinite string-localized \cite{M-S-Y}\cite{M-S-Y2} in the
sense of modular localization (which is the only relativistic quantum
localization). Since, as already remarked before, such fields cannot be
``interpolating'' and their S-matrix could not even be crossing symmetric, to
contemplate such a possibility would only make sense if the string field
theory S-matrix turnes out to really  violate crossing. 

In this context it is interesting to mention that recent results on string
localization lead to the apparent paradoxical conclusion that quantum
(modular) string localization does not admit a Lagrangian quantization
representation and classical Lagrangian string theories (e.g. Nambu-Goto) do
not lead to quantum string-localized objects. The coalescence of these two
different notions of localizations via quantization was a lucky circumstance
without which Pascual Jordan could not have succeded with his ``Quantelung der
Wellenfelder'' and QFT would have begun with the 1939 representation
theoretical approach of Eugene P. Wigner.

Whereas the construction of wedge algebras and their intersections based on
PFG particle properties seems to be limited to factorizing models, the idea of
getting to ambient wedge algebras and their intersections via ALH is
completely general. The lightfront algebras turn out to be transverse quantum
mechanical extensions of chiral QFTs and their classification does not appear
to be much more difficult than that of standard chiral theories on which a lot
of progress has been made. Among the ideas to construct QFTs in an intrinsic
manner, I consider the holographic projection setting the most promising.
Compared to the canonical ETCR setting it places the kinematics/dynamics cut
in such a way that the kinematical side (chiral theories) becomes much richer
and the dynamical side amounts to the reconstruction of the ambient theory
(inverse holography). This resembles in some way the role which chiral
theories are supposed to play in the dynamics of string theory. 

Among the many unsolved conceptional problems of QFT there is the question of
how particle-based concepts (S-matrix, formfactors crossing..) and the
causality based algebraic lightfront holography (transverse extended chiral
theories) fit together, e.g. questions like what is the holographic
interpretation of the $S_{scat}$ matrix? This is basically the old question
concerning the particle-field relation in a new context. 

The very fact that there are fundamental unanswered problems suggests that
despite its almost 80 years of existence, QFT still remains a project and is
still quite a distance from having reached maturity. It has remained young in
the sense of not having accomplished an ultimate formulation in purely
intrinsic terms, without the use of quasiclassical crutches with which Pascual
Jordan introduced field quantization, but from which he wanted to get
away\cite{Jordan}.

\textit{Acknowledgement}: I am indebted to Hradj Babujian and Michael Karowski
for interesting discussions extending over many years, and I owe special
thanks to J\"{o}rg Teschner for drawing my attention to the work of S.
Lukyanov and explaining some of its content.

\end{document}